\documentclass[aps,prd,superscriptaddress,nofootinbib,floatfix,twocolumn]{revtex4-2}

\usepackage{graphicx}
\usepackage{bm}
\usepackage{comment}
\usepackage{lastpage}  

\usepackage{enumitem}
\usepackage{amsmath} 

\newcommand{\hiroshima}{Graduate School of Advanced Science and Engineering, Hiroshima University, Kagamiyama, Higashi-Hiroshima, Hiroshima 739-8526, Japan}
\newcommand{\QUP}{International Center for Quantum-field Measurement Systems
for Studies of the Universe and Particles (QUP), KEK, Tsukuba, Ibaraki 305-0801, Japan}

\newcommand{\icr}{Institute for Chemical Research, Kyoto University Uji, Kyoto 611-0011, Japan}
\newcommand{\tokai}{Research Institute of Science and Technology, Tokai University, 4-1-1 Kitakaname, Hiratsuka, Kanagawa 259-1292, Japan}

\newcommand{\beq}{\begin{equation}}
\newcommand{\eeq}{\end{equation}}
\newcommand{\beqa}{\begin{eqnarray}}
\newcommand{\eeqa}{\end{eqnarray}}
\newcommand{\bpr}{\begin{problem}}
\newcommand{\epr}{\end{problem}}
\newcommand{\bcent}{\begin{center}}
\newcommand{\ecent}{\end{center}}
\newcommand{\bfig}{\begin{figure}}
\newcommand{\efig}{\end{figure}}
\newcommand{\bpc}{\begin{picture}}
\newcommand{\epc}{\end{picture}}
\newcommand{\barr}{\begin{array}}
\newcommand{\earr}{\end{array}}
\newcommand{\bitm}{\begin{itemize}}
\newcommand{\eitm}{\end{itemize}}
\newcommand{\bright}{\begin{flushright}}
\newcommand{\eright}{\end{flushright}}
\newcommand{\bminip}{\begin{minipage}}
\newcommand{\eminip}{\end{minipage}}
\newcommand{\btab}{\begin{tabular}}
\newcommand{\etab}{\end{tabular}}

\newcommand{\ro}{\rho}

\begin{document}
\title{Single-Point Search for eV-scale Axion-like particles 
with Variable-Angle Three-Beam Stimulated Resonant Photon Collider}

\author{Takumi Hasada}\affiliation{\hiroshima}
\author{Kensuke Homma*}\affiliation{\hiroshima}\affiliation{\QUP}
\author{Airi Kodama}\affiliation{\hiroshima}
\author{Haruhiko Nishizaki}\affiliation{\hiroshima}
\author{Yuri Kirita}\affiliation{\icr}
\author{Shin-ichiro Masuno}\affiliation{\icr}
\author{Shigeki Tokita}\affiliation{\icr}
\author{Masaki Hashida}\affiliation{\icr}\affiliation{\tokai}
\collaboration{${}^{\mathrm t}$SAPPHIRES collaboration}

\begingroup
\renewcommand\thefootnote{\fnsymbol{footnote}} 
\footnotetext[1]{Spokesperson and corresponding author}
\endgroup

\date{January 1, 2026}

\begin{abstract}
We report a laboratory search for axion-like particles (ALPs) in the eV-mass range using a variable-angle three-beam stimulated resonant photon collider. The scheme independently focuses and collides three laser beams, providing a cosmology- and astrophysics-independent test. By varying the angles of incidence, the center-of-mass energy can be scanned continuously across the eV range. In this work, we operated the collider in a vacuum chamber at a large-angle configuration, verified the spacetime overlap of the three short pulses, and performed a first search centered at $m_a\simeq 2.27~\mathrm{eV}$. No excess was observed. We thus set a $95\%$ C.L.\ upper limit on the pseudoscalar two-photon coupling, with a minimum sensitivity of $g/M\simeq 4.2\times 10^{-10}~\mathrm{GeV}^{-1}$ at $m_a=2.27~\mathrm{eV}$. This provides the first model-independent upper limit on the coupling that reaches the KSVZ benchmark in the eV regime and demonstrates the feasibility of eV-scale mass scans in the near future.
\end{abstract}

\maketitle

\section{Introduction}\label{Sec1}

 The existence of dark matter is indispensable for understanding the formation and evolution of cosmic structures. 
Measurements of the cosmic microwave background indicate that dark matter accounts for approximately $\Omega_{\mathrm{DM}}$ $\sim$ 0.27 of the total energy density of the Universe \cite{Planck}, yet its fundamental nature remains unknown. 

In particle physics, the axion \cite{AXION1,AXION2} and axion-like particles (ALPs) are well-motivated dark-matter candidates, arising as pseudo Nambu-Goldstone bosons of spontaneously broken global symmetries \cite{NGB1,NGB2,NGB3}. In the ALP {\it miracle} scenario, the portion of the inflaton condensate that survives the post-inflationary reheating phase can account for the present dark-matter abundance, predicting an ALP (inflaton) mass in the range 0.01 $\sim$ 7.7 eV \cite{MIRACLE1,MIRACLE2}. On the other hand, for cold-hot dark-matter scenarios, it is discussed that a cold population of light bosons can be produced directly from the primordial thermal bath. Bose enhancement rapidly 'bursts' low-momentum modes, generating a light boson around eV mass scale \cite{CHOT1,CHOT2}.

Also, phenomenologically there are indications favoring eV-scale ALPs. Recent measurements of the cosmic optical and infrared background (COB/CIB) have suggested possible excesses of diffuse emission in some bands. One intriguing interpretation attributes this excesses to two-photon decays of ALPs\cite{LORRI1,HST,LORRI2}. Independently, measurements of optical-depth of $\gamma$-rays from blazars show slightly stronger attenuation than that predicted by galaxy-count-based EBL models. Modeling the residual optical depth as photons from ALP two-photon decays favors eV-scale ALPs \cite{gamma}. In the eV-mass region, several indirect-detection searches have been carried out \cite{JWST,MUSE,VIMOS,WINERED}. Among them, WINERED experiment used high-resolution spectroscopy to probe the $1.8~\text{-}~2.7~\mathrm{eV}$ window, setting stringent limits, and while the observing conditions were not ideal, several local $>5\sigma$ features were noted \cite{WINERED}.

\begin{figure}[!htbp]
         \centering
         \includegraphics[width=0.52\textwidth]{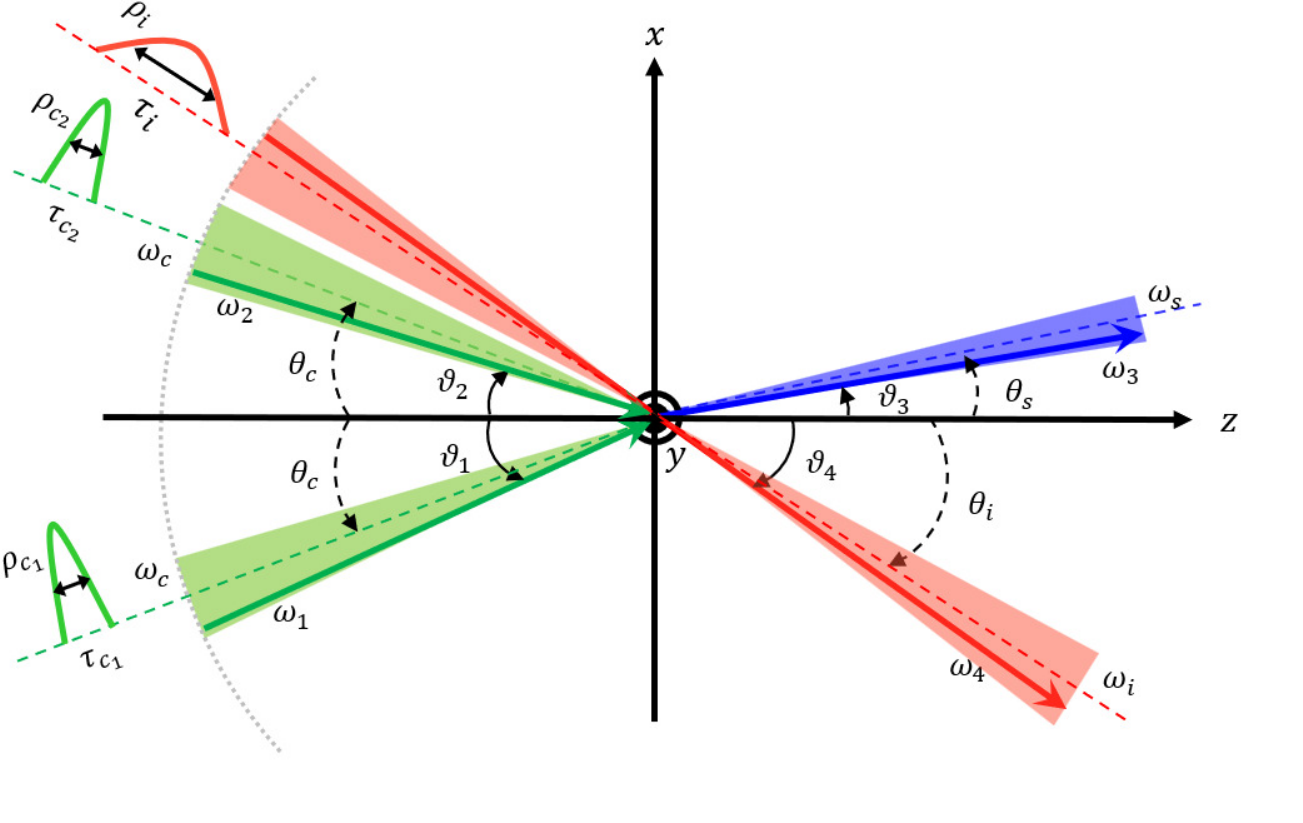}
         \caption{
Concept of the three-beam stimulated resonant photon collider (${}^{\mathrm{t}}$SRPC). Three short-pulsed laser beams with photon number densities $\ro_k (k=c_1, c_2, i)$ are independently focused at the collision point. The two creation beams (green) cross at an incidence angle $\theta_c$ and produce an ALP resonance, while the inducing beam (red) stimulates its decay. The creation photons have energies $\omega_1$ and $\omega_2$ and incidence angles $\theta_1$ and $\theta_2$ around the central values $\omega_c$ and $\theta_c$, respectively. Similarly, the inducing beam with central energy $\omega_i$ contains a frequency component $\omega_4$ that enhances emission of signal photons with energy $\omega_3$ (blue) via energy-momentum conservation.}
     \label{Schematic}
\end{figure}

In this paper, as a laboratory-based search in the eV-mass scale, we present the result of the search based on stimulated resonant photon collisions (SRPC) \cite{PTEP2014,PTEP2015,PTEP2020,UNIVERSE,JHEP} using three-laser beams. 
Figure \ref{Schematic} shows a schematic view of the three-beam stimulated resonant photon collider (${}^{\mathrm{t}}$SRPC).
This method enables model-independent testing, free from astrophysical, astronomical and cosmological assumptions, by colliding two focused creation laser beams with a third focused inducing laser beam. The pair of creation lasers generates ALP-resonance states at the collision point. With the average energy of a single photon in the creation lasers, denoted by $\omega_c$, and subtended angles $2\theta_c$ between colliding two photons, the center-of-mass system energy is given by
\begin{equation}
E_{cms}=2\omega_c\sin\theta_c.
\end{equation}
The inducing laser stimulates the decay of the generated ALP. Denoting the average single-photon energy in the inducing field as $\omega_i$, one of the two decay photons can be extracted as the signal photon with energy $\omega_s=2\omega_c-\omega_i$, which is kinematically equivalent to the atomic four-wave-mixing (FWM) process.
In practice, photons from each beam collide with a spread of angles and energies: focusing induces transverse-momentum (angular) spread, and the ultrashort pulse duration gives a finite energy bandwidth. As a result, the center-of-mass system energy is effectively modified to
\begin{equation}
E_{cms}=2 \sqrt{\omega_1\omega_2}\sin\left( \frac{\vartheta_{1} + \vartheta_{2}}{2} \right),
\end{equation}
and a fraction of the inducing field $\omega_4$ stimulates scattering into signal photons of energy $\omega_3$ that satisfy energy-momentum conservation. Consequently, the collision energy is not fixed but distributed, and by changing the incidence angle $\theta_c$ with a step matched to the mass resolution, we can perform a fast and continuous mass scan.
We have therefore proposed \cite{3beam00}, peformed the pilot search \cite{3beam01} and developed a variable-incidence three-beam collider that enables a continuous eV-scale scan \cite{3beam02} (see also sub-eV searches in quasi-collinear, coaxially focused two-beam configurations performed by the SAPPHIRES Collaboration \cite{SAPPHIRES00,SAPPHIRES01,SAPPHIRES02}).

In our previous studies, we established a method to ensure precise spatiotemporal overlap at the focal point of the three-laser collision and performed searches using low-energy laser fields under atmospheric conditions \cite{3beam01}. Furthermore, we constructed a variable-angle three-beam stimulated resonant photon collider and verified its operational feasibility \cite{3beam02}. In this paper, we present the result of the first point search with the variable-angle collider in the vacuum. The search was performed at the large angle of incidence between two creation beams, $\theta_c \pm 47.9^\circ$, corresponding to ALP mass $m_a \sim 2.27~\mathrm{eV}$, which is the most difficult angle to adjust for achieving the three-beam spatiotemporal overlap. The energies of the creation lasers and the inducing laser were each increased by more than three orders of magnitude compared to the previous search at the atmosphere.

This paper is organized as follows. we first describe the experimental setup designed for the search under vacuum conditions. Secondly, we detail the measurement. Thirdly, we present the analysis results and the null result. Finally, we provide the new upper limit based on the set of the experimental parameters and conclude the search.

\section{Experimental setup}\label{Sec2} %

\begin{figure*}[t]
         \centering
         \includegraphics[width=1.00\textwidth]{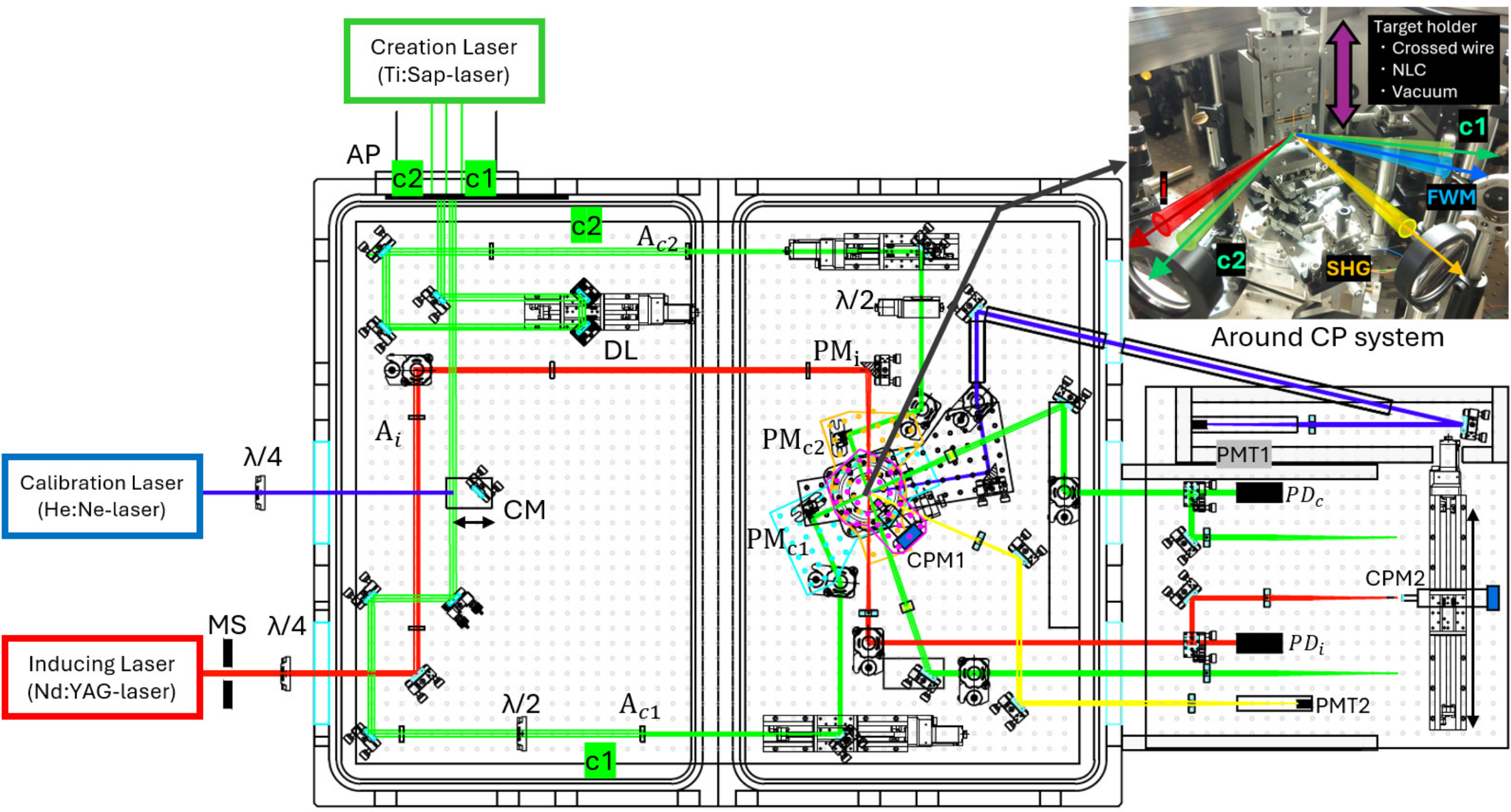}
         \caption{
Schematic of the search setup.
A Ti:Sapphire laser (green, $\lambda_c=811\,\mathrm{nm}$) enters from the upper left and serves as the creation laser, while an Nd:YAG laser (red, $\lambda_i=1064\,\mathrm{nm}$) enters from the left and serves as the inducing laser. Both are brought to intersect at the collision point (CP) inside a vacuum vessel, where axion-like particles (ALPs) can be resonantly produced and induced to decay.
\textbf{Creation path (green):} Upstream, the beam is clipped by a aperture plete (AP) into two creation beams, c1 and c2. Each is then cut to a diameter of $5\,\mathrm{mm}$ by apertures $A_{c1}$ and $A_{c2}$. A delay line (DL) on the c2 arm tunes the optical path length. In each optical path, a 1.81-mm-thick half-wave plate ($\lambda/2$) is inserted to make the polarization states orthogonal at the CP. The beams are directed to the variable-angle three-beam stimulated resonant photon collider on the right side of the vacuum chamber and focused to the CP by parabolic mirrors ($PM_{c1},PM_{c2}$) at incidence angles $\theta_{c1,c2}=\pm 47.9^\circ$ to produce an ALP resonance. And after collimation by a lens, c1 line is measured by a photodetector $PD_c$, and the laser profile c1, c2 at CP is monitored by a collision-point monitor (CPM2).
\textbf{Inducing path (red): }
Upstream, a mechanical shutter (MS) generates a special trigger pattern. The beam is propagated in a circular polarization state by quarter plate $\lambda/4$ and focused onto the CP by a parabolic mirror ($PM_i$) at an incidence angle $\theta_i=-66.4^\circ$ to induce the decay of the resonant state. After collimation by a lens, inducing laser is recorded by $PD_i$, and the CP profile is verified with CPM2.
\textbf{Variable-angle three-beam stimulated resonant photon collider:}
The collider is a four-layer assembly of rotation stages and aluminum plates. To independently vary collision angles and focus for each laser beam, parabolic mirrors are mounted on an aluminum plate. The lowest (black) signal layer collimates the ALP-mediated signal photons (blue:~$\sim 655\,\mathrm{nm}$) emerging from the CP with a parabolic mirror and guides them to photomultiplier-tube (PMT1). The middle layers (cyan for c1, orange for c2) focus the creation beams. The top (magenta) layer hosts a collision-point monitor (CPM1) for alignment of the laser profiles and the incidence angles; this camera is removed during data taking. At the collider center, a target system is set to ensure spatiotemporal overlap of all three beams and to enable automated vertical-axis switching among a cross-shaped wire, a nonlinear crystal (NLC), and vacuum.
\textbf{Calibration and signal path (blue):}
To measure the acceptance of PMT1 for photons originating at the CP, a He-Ne laser $633\,\mathrm{nm}$ is injected by temporarily inserting a calibration mirror (CM) and routing the beam along the c1 line from the left. During the acceptance measurement, the c1 layer (cyan) is temporarily aligned to match the signal layer's (black) divergence $\sim 34.0^\circ$ from the CP, reproducing the signal-photon path to PMT1.
 }
     \label{setup}
\end{figure*}

Figure \ref{setup} shows a schematic view of the search setup in the large-angle configuration. From the left side of the figure, the creation laser (indicated in green) and the inducing laser (indicated in red) are introduced. We used a Ti:Sapphire laser (T6-system) with a pulse duration of $\sim$~34 fs for the creation field and a Nd:YAG laser with a pulse duration of $\sim$~7.1 ns for the inducing field. Both lasers are available at the Institute for Chemical Research, Kyoto University. The central wavelengths of these lasers were 811 nm and 1064 nm, respectively. In the right side of figure, the variable-angle collider is configured at the large incidence angle. The collider comprises four layers combining rotation stages with aluminum plates, allowing independent control of the incidence angles for three beams (details in Ref.~\cite{3beam02}). The angles of incidence were set to $\theta_c = \pm 47.9^\circ$ for the creation beams, and $\theta_i = -66.4^\circ$ for the inducing beam with respect to the bisecting angle between the two creation laser beams. Consequently, the resonance mass determined by the central wavelengths and the central angle of incidence corresponds to $m_a \sim 2.27$ eV. From energy and momentum conservation, the signal photon is expected to have a wavelength $\omega_s = 655\ \text{nm}$ and an emission angle of $\theta_s \sim 34.0^\circ$.

The creation laser beam, with an initial diameter of approximately $\sim40$ mm, is split into two paths, c1 and c2, using an aperture plate (AP) with two $\sim$ 10 mm diameter apertures. The beam diameters of the three laser lines are defined by apertures($A_{c1},A_{c2},A_i$) to be 5 mm for c1 and c2, and 7 mm for the inducing beam path, i. Each beam is independently focused by a parabolic mirror ($PM_{c1},PM_{c2},PM_{i}$) in the collider. The focal lengths are $f_c = 101.6$ mm for the c1 and c2 beams and $f_i = 203.2$ mm for the inducing beam. The optical paths of the creation lasers (c1 and c2) each include a 1.81-mm-thick half-wave plate ($\lambda$/2), and the polarizations are made mutually orthogonal at the focus to ensure sensitivity to a pseudoscalar field. The principle of the beam polarization measurement are given in Ref.~\cite{3beam01}. In i-path for the inducing laser, a quarter-wave plate ($\lambda$/4) is inserted, and the polarization is set to right-handed circular as viewed from an observer at the collision point (CP). A mechanical shutter is placed at i-path and is used to generate the special trigger pattern described in the next section.

The spatiotemporal overlap of the three colliding laser beams is ensured by a specialized target system placed at the focus, with collision point monitors (CPM1, CPM2) and an array of photodetectors ($PD_{c}, PD_i$). CPM1, mounted on a motorized rotation stage with magenta layer, is used to check the focal position of each laser beam and to set the incidence angle with an accuracy of $\sim 0.1^\circ$. It is removed for the search runs in the vacuum. CPM2, mounted on an x-axis translation stage, is optically conjugate to the CP and enables three-beam focal-profile checks with a single camera. A target system is equipped with a cross-shaped wire of 10 $\mu$m diameter, a nonlinear crystal (NLC), and a “no target” option (vacuum), all of which can be precisely switched in the vertical positions using a motorized stage. 

\begin{figure*}[t]
        \centering
        \includegraphics[width=1.00\textwidth]{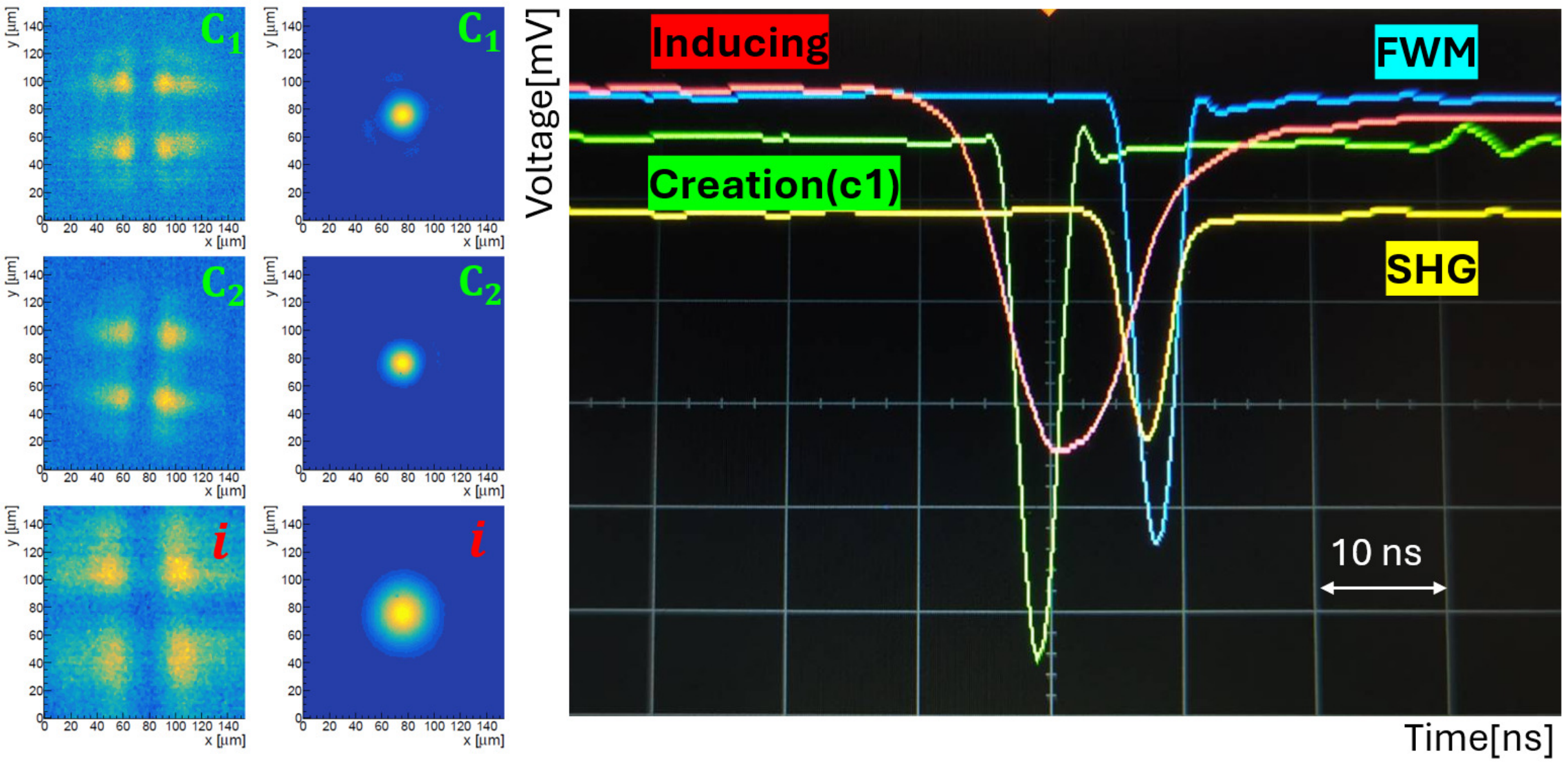}
        \caption{
\textbf{Left, middle:} Beam-profile images of c1, c2, and i recorded with the collision-point monitor (CPM2). In left three panels, a $10\,\mu\mathrm{m}$-diameter cross-shaped wire is placed at the collision point (CP) as a common target. The spatial overlap of the three laser beams is ensured by imaging onto the wire at the focal point and aligned to coincide. Along the focusing direction ($z$-axis), the image is formed so that the vertical wire is in focus. For each beam, single camera pixel corresponding to the center of the cross-shaped wire is defined. During the search, the cross-wire is removed and each beam's profile centroid is aligned to the deﬁned camera pixel at the search-mode waist as shown in middle three pannels.
\textbf{Right:} Oscilloscope trace of four-wave mixing (FWM). A nonlinear crystal is installed at the CP under vacuum to ensure three-beam spatio-temporal overlap. Scanning the delay line shows that second-harmonic generation (SHG, yellow) occurs only when the two creation beams (c1, c2, green) overlap at the CP, whereas FWM (cyan) appears only when all three beams-including the inducing beam i overlap.
}
    \label{overlap}
\end{figure*}

The left and middle panels in figure \ref{overlap} show a set of focal images of the three laser beams at CP, monitored by the CPM2. The spatial overlap of the three laser beams is ensured by imaging them onto the cross-shaped wire placed at the focal point and adjusting them to coincide relative to one another as showed in the most left three pannels. Along the beam focusing direction (the $z$ axis), the focal point is adjusted so that the vertical wire is in focus. Once the alignment with the wire is completed, the central position of the cross-shaped wire is mapped upon one camera pixel. By aligning the center of each laser beam to this defined pixel, spatial overlap can be consistently maintained as shown in middle three pannels obtained without wire-target.
Right panel of figure \ref{overlap} shows an oscilloscope image captured when all the three laser pulses overlap. The temporal overlap at CP is verified using nonlinear optical phenomena via the NLC. A delay line (DL) composed of a retroreflector is installed at c2-path to allow fine adjustment of the optical path length. Second harmonic generation (SHG:$2\omega_c \sim 405$~nm) indicated in yellow is generated only when the two green creation laser pulses overlap. By scanning the DL to maximize the SHG signal on a photomultiplier tube (PMT2), we ensure the temporal overlap of the two creation lasers. The overlap of three beams is ensured by observing a four-wave mixing (FWM) by a photomultiplier for the signal photon detection (PMT1). The red inducing laser is triggered by the creation laser system to ensure synchronization via electrical signals. FWM (cyan) is observed only when all three lasers are synchronized. In order to suppress noise components, optical filters are placed in front of each photomultiplier tube (PMT1, PMT2): one centered at 400 $\pm$ 5 nm to select SHG and another at 660 $\pm$ 10 nm to select FWM.

The NLC is mounted on a rotation stage with its surface perpendicular to the angle bisector of the two green creation beams.Taking the NLC surface as the CP for all three beams minimizes refractive-index-driven spatiotemporal offsets relative to the vacuum-target configuration to negligible levels.

A He:Ne laser (blue), with a wavelength of 633 nm chosen to be close to that of the expected signal photon is introduced from the left side of the setup for the signal acceptance measurement. During this measurement, inserting a calibration mirror (CM), c1-path is temporarily used to mimic the signal path from CP, ensuring the correct signal path. The laser fluxes are measured at two points at CP and at the position of PMT1 using the same camera and the flux ratio gives the geometrical acceptance factor for signal photons. During acceptance measurements, the two bandpass filters: a long-pass filter transmitting wavelengths above 650 nm and a short-pass filter transmitting below 670 nm are removed. The acceptance measurement using the 633 nm He:Ne laser is conducted under a filtering condition that transmits the $570~\text{-}~690$ nm wavelength range.

\section{Measurement} \label{Sec3}

\begin{figure}[!htbp]
        \centering
        \includegraphics[width=0.48\textwidth]{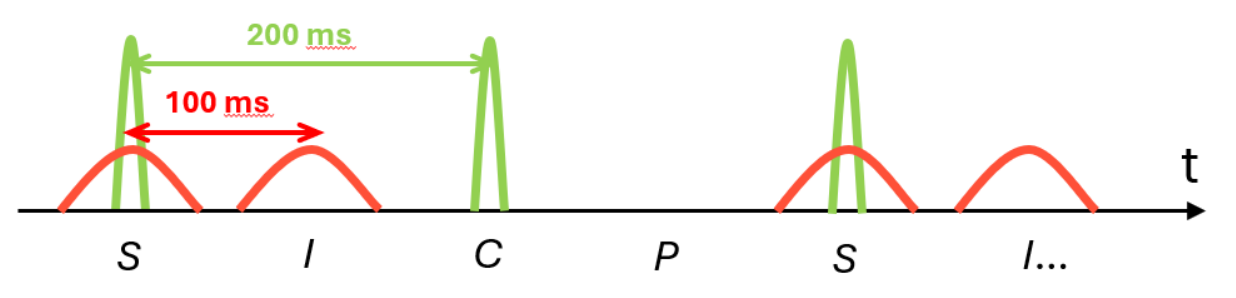}
        \caption{
Trigger pattern used during the search.
A four-state trigger sequence is introduced by combining the emission states of the three beams to form \emph{S}, \emph{I}, \emph{C}, and \emph{P} patterns. The classifications are:
S for both laser pulses, C for only the creation laser pulses, I for only the inducing laser pulses, 
and P for pedestals without laser pulses.
}
    \label{trigger}
\end{figure}

The creation lasers (the T6-laser system) was operated at a repetition rate of \(5\,\mathrm{Hz}\), whereas the inducing laser (Nd:YAG) was operated at \(10\,\mathrm{Hz}\).
For the ALP search, a mechanical shutter was inserted into the inducing laser beamline to form the trigger patterns in Fig.~\ref{trigger}.

The laser composition in each trigger pattern is:
\begin{enumerate}
  \item \textbf{S-pattern:} creation lasers plus inducing laser.
  \item \textbf{C-pattern:} only creation lasers.
  \item \textbf{I-pattern:} only inducing laser.
  \item \textbf{P-pattern:} no laser (pedestal).
\end{enumerate}

For the ALP search, we evaluated the number of photons observed by PMT1. 
Let us define \(N_{S},N_{C},N_{I},N_{P}\) as the observed numbers of photons per shot for the \(S,C,I,P\) patterns, respectively, 
by assuming the following linear superpositions:

\begin{eqnarray}
N_{S} &=& n_{\mathrm{sig}} + n_{c} + n_{i} + n_{p}, \quad N_{C} = n_{c} + n_{p}, \\ \nonumber
N_{I} &=& n_{i} + n_{p},\qquad \qquad \qquad N_{P} = n_{p}.
\end{eqnarray}
The pedestal \(n_{p}\) is present in all the patterns, whereas the noise photons originating from the creation and inducing lasers, \(n_{c}\) and \(n_{i}\), appear in their respective single-beam C, I patterns and also in the \(S\) pattern. The observed number of four-wave mixing (FWM) photons \(n_{\mathrm{sig}}\) arises only in the \(S\) pattern.
%
The number of signal photons can be extracted as
\begin{equation}
\label{signal_logic}
n_{\mathrm{sig}} = N_{S} - N_{C} - N_{I} + N_{P}\
\end{equation}
by taking the linear combinations of the number of photons observed in the four trigger patterns.

\begin{figure}[!htbp]
        \centering
        \includegraphics[width=0.45\textwidth]{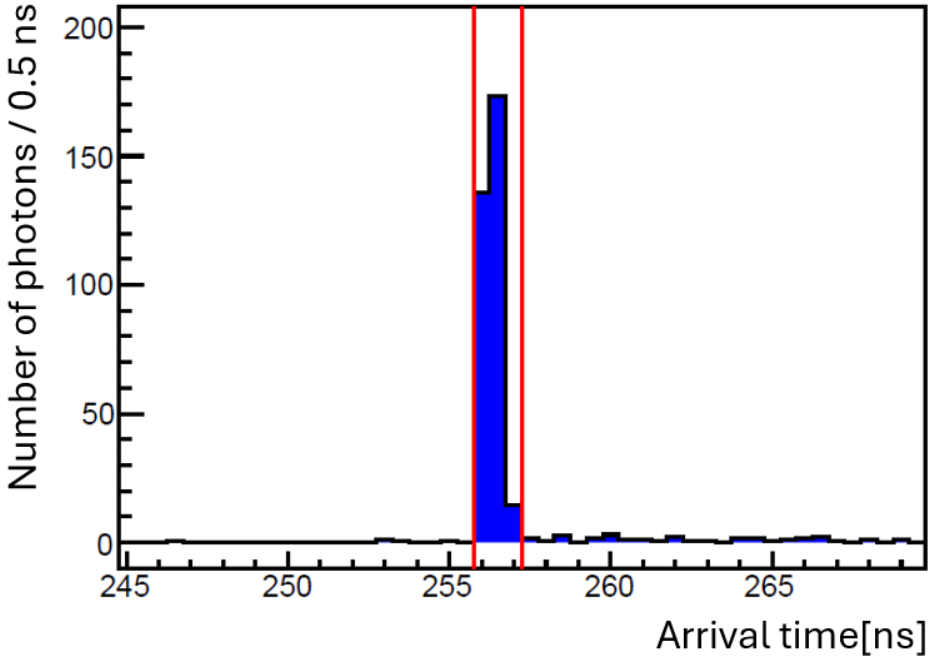}
        \caption{
Arrival time distribution of FWM photons for the S pattern (500 shots) via the atomic process 
when the NLC was placed at CP.
The red lines thus provide the expected time window for FWM photons via ALP-exchange to arrive. 
}
        \label{BBOfwm}
\end{figure}
        
Figure \ref{BBOfwm} shows the arrival time distribution of FWM-like photons $N_{S}$ for the S pattern (500 shots) measured using PMT1 with NLC. The number of photons was reconstructed from the voltage-time relation of analog signals from PMT1 with a waveform digitizer. We also applied a peak-finding algorithm to simultaneously determine the number of photons and their arrival times from falling edges of amplitudes in the waveforms. The details of the peak-finding algorithm are given in the Sec. \ref{Sec4}. The horizontal axis represents the photon arrival time with a bin width of $\Delta t=0.5$ ns, and the vertical axis shows the number of reconstructed photons per 0.5 ns.
The arrival time of the FWM photons originating from the atomic process in NLC was used to define the time window 
corresponding to that for signal photon emission via ALP exchanges. With the creation and inducing lasers operated at low energy, the yields \(N_{C}\) and \(N_{I}\) in the C and I patterns were negligible. In the search experiment, we evaluated the number of photons within this red time window for counting the number of ALP-induced signal photons.

During the search, three beam focal spots can drift due to thermal variations in the laser system. To suppress the drift in focal position over time and to ensure good overlaps, we conducted data taking 20,000 shots (5,000 shots per trigger pattern: S, I, C, P) per run and performed three runs in total. Before and after individual runs, the focal positions of the laser beams were monitored with the CPM2 to assess the spatiotemporal misalignment. The measured drifts were used to evaluate the averaged three-beam
overlaping factor as explained later.

\section{Data analysis} \label{Sec4}

\begin{figure}[!htbp]
        \centering
        \includegraphics[width=0.50\textwidth,height=0.45\textheight,keepaspectratio]{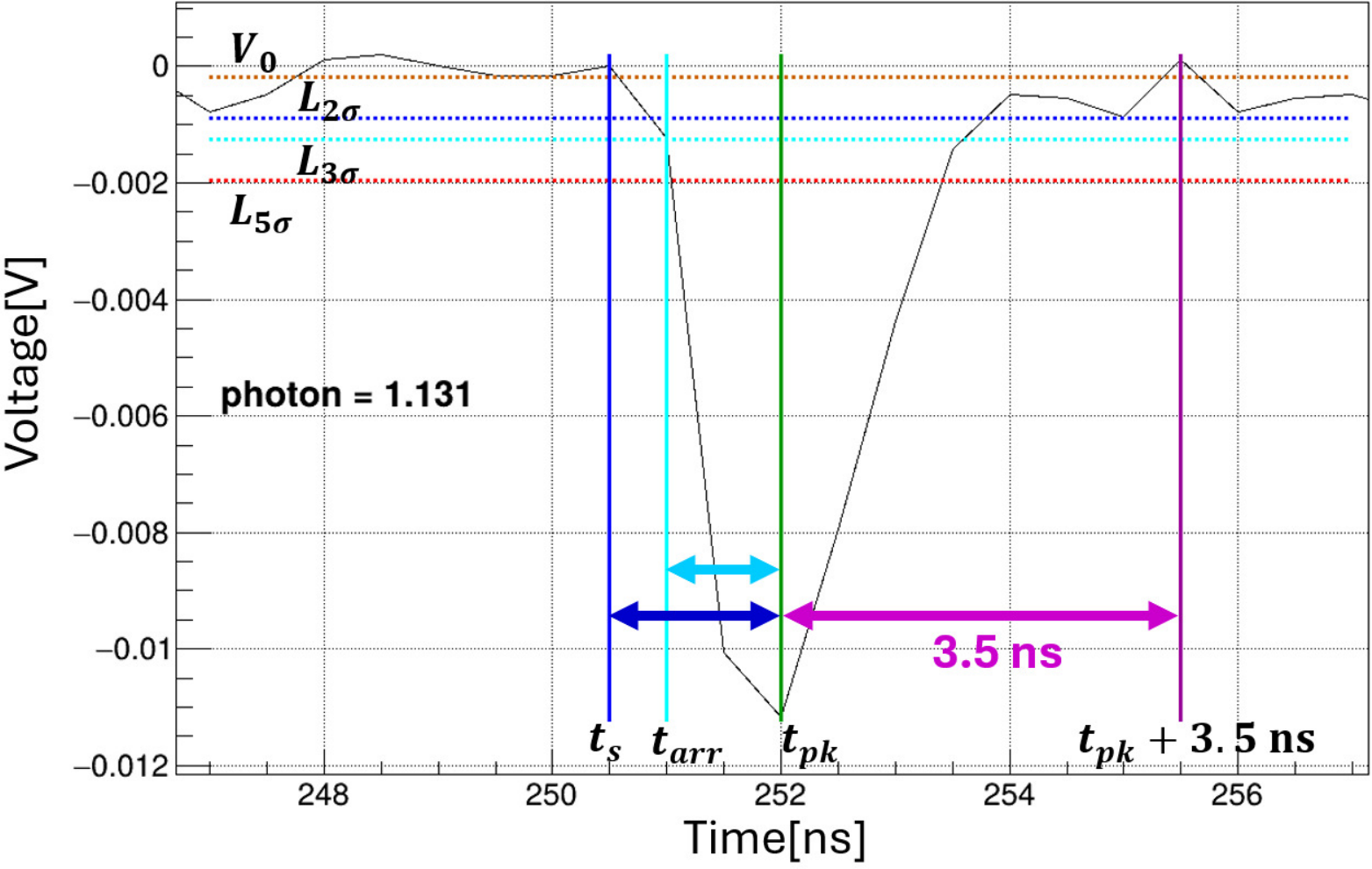}
        \caption{
Typical waveform and peak-finding algorithm measured by the digitizer. The orange dashed horizontal line shows the baseline mean $V_{0}$, estimated over the initial baseline window \(t\in[0,100\,\mathrm{ns}]\). The red dashed horizontal line is the photon-detection threshold \(L_{5\sigma}\). The green vertical line marks the peak time \(t_{\mathrm{pk}}\) (time of the most negative sample that exceeds \(L_{5\sigma}\)). The cyan and blue dashed horizontal lines are the arrival-time threshold \(L_{3\sigma}\) and the integration-start threshold \(L_{2\sigma}\), respectively; these define the arrival time \(t_{\mathrm{arr}}\) (cyan vertical line) and the integration-start time \(t_{\mathrm{s}}\) (blue vertical line), respectively. The purple vertical line indicates the integration-end time \(t_{\mathrm{e}}\), set to 3.5 ns after the peak time. The waveform charge is obtained by integrating from \(t_{\mathrm{s}}\) to \(t_{\mathrm{e}}\), then converted to photons using one-photon-equivalent charge from the PMT calibration (details in Appendix \ref{App.A}). In this example, the estimated photon count is $\sim$ 1.1.
}
\label{PeakFinding}
\end{figure}

We evaluated the signal-photon yield with a PMT (R9880U-20) read out by a digital oscilloscope (Acqiris U1065A). A representative zoomed waveform is shown in Fig.~\ref{PeakFinding}. The vertical axis is digitized with \(10\)-bit resolution relative to the full-scale input voltage, and the horizontal axis covers \(500\,\mathrm{ns}\) (full scale), sampled at period \(\Delta t=0.5\,\mathrm{ns}\).
For the analysis, we employ a peak-finding algorithm for two reasons:
(1) to separate baseline electrical noise in the digital oscilloscope from photon-induced pulses; and
(2) to use pulse arrival time to discriminate background/ambient light from the time-correlated signal.
We implement the peak-finding algorithm as follows:

\begin{enumerate}[label=(\roman*)]
\item \textbf{Baseline and threshold.}
Let \(V[t]\) denote the digitized PMT waveform (in volts) per shot as a function of time \(t\). For each shot, define the initial baseline window as \(t\in[0,100\,\mathrm{ns}]\). From the ensemble of baseline voltages over all shots, fit a Gaussian to obtain the baseline RMS $\sigma_v$. For each shot, compute the baseline mean $V_{0}$ and define the per-shot threshold level
\begin{equation} \label{ThrPhotonpeak}
L_{5\sigma} \equiv V_{0}-5\sigma_v \, .
\end{equation}

\item \textbf{Peak Identification.}
For each shot, count pulses that exceed $L_{5\sigma}$. For each pulse, identify the \emph{peak time} \(t_{\mathrm{pk}}\) as the most negative sample (the minimum voltage) within that pulse.

\item \textbf{Arrival time and integration range.}
Define two voltage threshold levels per shot:
\begin{align}
L_{3\sigma} &\equiv V_{0} -3\sigma_v, \label{ThrArrive}\\
L_{2\sigma} &\equiv V_{0} -2\sigma_v. \label{ThrStart}
\end{align}
Scanning backward in time from \(t_{\mathrm{pk}}\), the \emph{arrival time} $t_{\mathrm{arr}}$ and \emph{integration-start time} $t_{s}$ are defined as the first upward crossings of $L_{3\sigma}$ and $L_{2\sigma}$, respectively.
%
%
The index \(t_{\mathrm{arr}}\) denotes the arrival time of pulse with threshold level at $L_{3\sigma}$ to mitigate false triggers from baseline fluctuations. The index \(t_{s}\) marks the start time of charge integration for pulse, using the milder threshold level $L_{2\sigma}$.

Set the \emph{integration-end time} $t_{e}$ to be $3.5\,\mathrm{ns}$ after peak time,
%
\begin{equation}
t_{e} \;=\; t_{\mathrm{pk}} + 3.5~\mathrm{ns},
\end{equation}
chosen to include sufficient charge and to avoid contamination from PMT afterpulses, which typically occurs at $\sim 5~\mathrm{ns}$ after the primary peak~\cite{pmt}.

%

\item \textbf{Charge assignment.}
Compute the baseline-subtracted time-integral voltage between $t_{s}$ and $t_{e}$ with
$\Delta t = 0.5~\mathrm{ns}$ as
\begin{equation}
Q 
\;=\;
\sum_{t_{s}}^{t_{e}}
\bigl( V[t]- V_{0} \bigr)\,\Delta t / R \, ,
\end{equation}
where $R = 50~\Omega$ is the nominal input termination of the digitizer; since $R$
cancels in the calibration ratio, its precise value does not affect the final results.
The resulting charge $Q$ is then assigned to the arrival timestamp $t_{\mathrm{arr}}$.

\item \textbf{Photon conversion.}
Using the single-photoelectron calibration of the PMT (Appendix~\ref{App.A}), performed with the same PMT, cabling, and front-end electronics as in the search data, we determine the single-photon-equivalent charge to be
\begin{equation}
Q_{1\mathrm{PE}}  = -0.2337 \pm 0.0004~[\mathrm{pC/photon}].
\label{Q1PE}
\end{equation}

The corresponding number of detected photons is
\begin{equation}
N \;=\; \frac{Q}{Q_{1\mathrm{PE}}}\,.
\end{equation}
\end{enumerate}
Applying steps (i)--(v) to every shot and the peak-finding algorithm enables time-resolved photon counting
within the time window where signal photons are expected to arrive.

\begin{figure*}[t]
        \centering
        \includegraphics[width=1.0\textwidth]{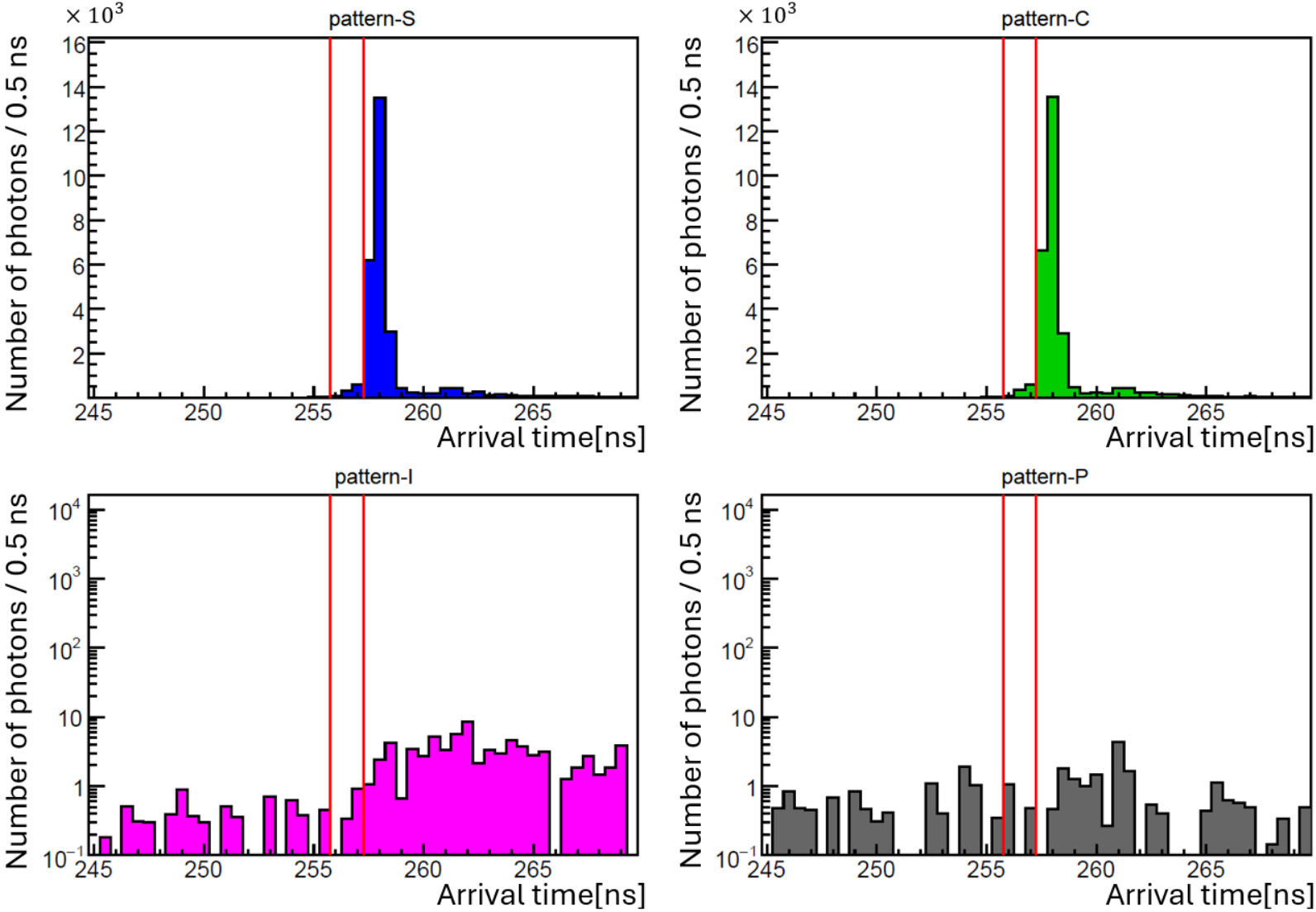}
        \caption{
Arrival-time distributions of detected photons with $15{,}000$ shots in each trigger pattern measured at $2.0\times 10^{-3}\,\mathrm{Pa}$. The upper-left, upper-right, lower-left, and lower-right histograms correspond to the distributions in S, C, I, and P trigger patterns, respectively. The \textsc{I} and \textsc{P} panels are plotted on a logarithmic scale. The red vertical lines represent the time window in which signal-like FWM photons are expected to arrive.
}
        \label{SCIP}
\end{figure*}

Figure \ref{SCIP} shows histograms in the upper left, upper right, lower left and lower right corresponding to S, C, I and P patterns of beam combinations, respectively. A total of \(60{,}000\) shots were collected across three runs, with \(15{,}000\) shots in each trigger pattern.
In both the S- and C-patterns, a peak distribution of photon counts was observed mainly after the designated signal time (red window). Since no pressure dependence of the photon yield was observed and the similar distribution was also seen in the C-pattern, this background is likely due to scattered lights from optical components encountered by the creation laser (c1, c2) during the propagation. Although identifying and suppressing this noise source remains a future task, the main noise distribution lies outside the arrival time window of the signal photons. We thus evaluate the number of signal photons based on Eq.~\eqref{signal_logic}. 

\begin{figure}[!htbp]
        \centering
        \includegraphics[width=0.45\textwidth]{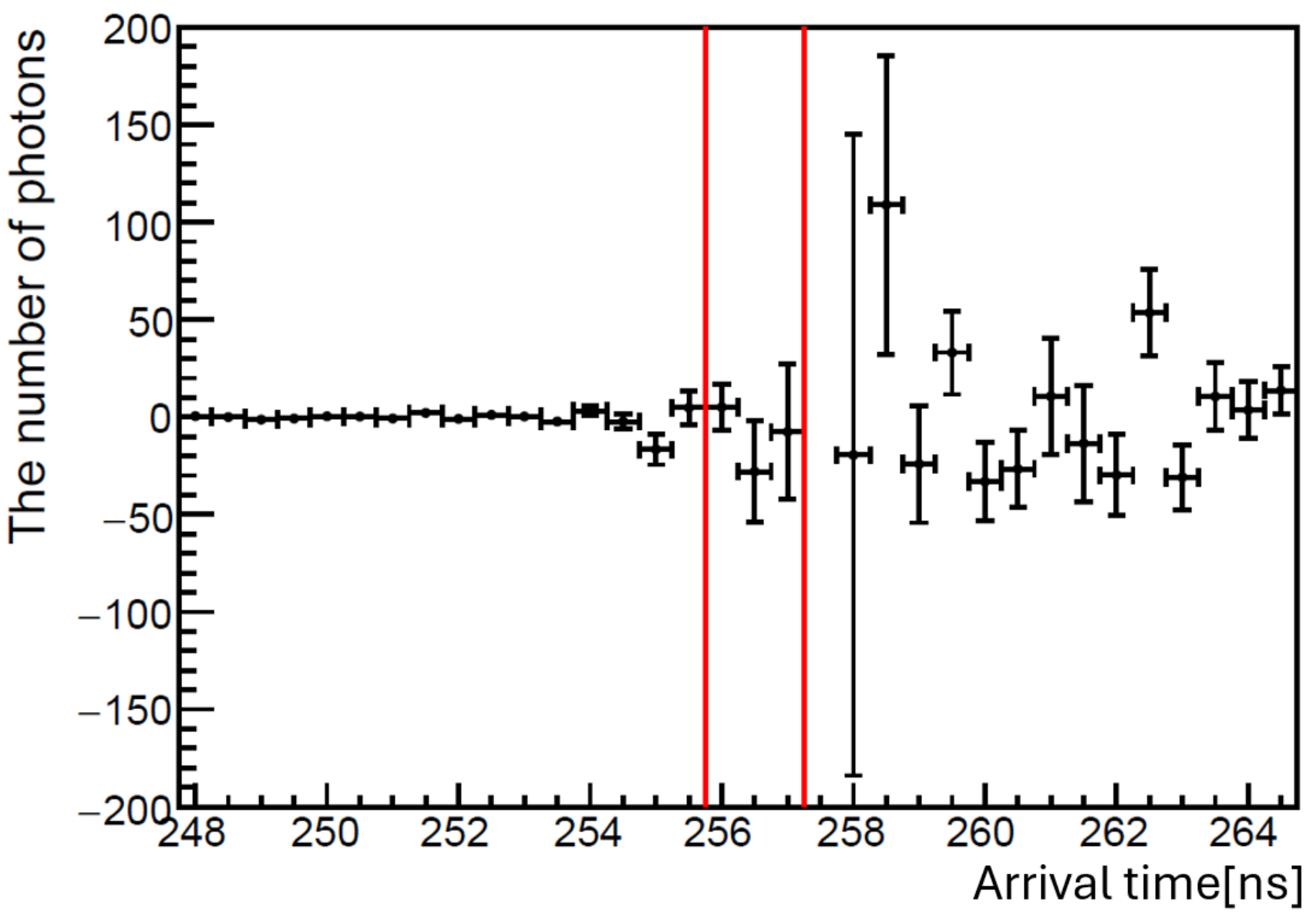}
        \caption{
Arrival-time distribution of detected photons after the $S-C-I+P$ subtraction. The red window marks the expected arrival-time gate for ALP-exchanged signal photons.
}
        \label{FWM}
\end{figure}

Figure \ref{FWM} shows the result after $S-C-I+P$ subtraction with the statistical errors. 
The number of photons observed within the red time window was obtained as
\begin{eqnarray}
\label{signal_photon}
n_{\mathrm{obs}} = -30.5
  \pm 52.4\,\text{(stat.)} 
  \pm 13.9\,\text{(syst.\,I)} \\ \nonumber 
  \pm 25.5\,\text{(syst.\,II)} 
  \pm 30.6\,\text{(syst.\,III)}, 
\end{eqnarray}
where the first uncertainty is statistical, evaluated from the Poisson fluctuations of the reconstructed photon numbers in the four trigger patterns, including the Fano-factor enhancement due to the Gamma-distributed single-photon charge response, as given by Eq.(\ref{stat}) and discussed in Appendix \ref{App.B}, while
the three systematic errors are coming from the definition of the peak-finding algorithm as follows.
The first systematic error (syst.I) was evaluated by varying the voltage threshold level in Eq.~\eqref{ThrPhotonpeak} used to discriminate photon-peaks from baseline fluctuations,
        \begin{equation}
        L_{5\sigma} = V_{0}-5\sigma_v\pm1\sigma_v.          \end{equation}
The second systematic uncertainty (syst.II) was estimated by changing the voltage level of arrival-time threshold in Eq.~\eqref{ThrArrive},
        \begin{equation}
         L_{3\sigma} = V_{0}-3\sigma_v\pm1\sigma_v. 
        \end{equation}
The third systematic uncertainty (syst.III) was evaluated by varying the voltage threshold level in Eq.~\eqref{ThrStart} that defines the initial bin of the charge-integration window,
        \begin{equation}
         L_{2\sigma} = V_{0}-2\sigma_v\pm1\sigma_v.
        \end{equation} 
The observed number of signal photons $n_{obs}$ was consistent with zero within the experimental uncertainty.

Other expected systematic errors arising from the degree of spatiotemporal overlap among the three beams during a single run will be 
discussed in the next section, where the effective overlap factors for individual runs will be evaluated and 
then incorporated to derive the upper limit from the null result.

\section{Upper limit in the coupling-mass relation for ALP-exchange} \label{Sec5}
As no significant signal was observed in Section~\ref{Sec4}, we derive an exclusion region in the ALP coupling-mass plane. This is done using the formulation of the signal-photon yield presented in~\cite{3beam00}, together with the experimentally determined total uncertainty, as follows.
The expected signal-photon yield per pulse collision in stimulated resonant scattering, $\mathcal{Y}_{c+i}$, is given by
        \begin{equation}
        \label{yield0}
        \mathcal{Y}_{c+i} \equiv N_{1} N_{2} N_{4} \mathcal{D}_{three} \left[s / L^3\right] \bar{\Sigma}_I \left[L^3 / s\right],
        \end{equation}
where $N_{1}(=N_{c1}), N_{2}(=N_{c2})$ and $N_{4}(=N_{i})$ are 
the average photon numbers in each laser, $\mathcal{D}_{three}$ represents the spatiotemporal overlap factor among the three focused beams at 
the collision~\cite{3beam00}, and $\bar{\Sigma}_I$ denotes the volume-wise interaction rate~\cite{JHEP,UNIVERSE}. The corresponding units are expressed in [~~], with $L$ for length and $s$ for time. 

The center–of–mass system energy \(E_{\mathrm{cms}}\) of two colliding laser photons is
intrinsically smeared by quantum uncertainty: each photon must be treated as a wave packet
with finite temporal duration, which implies a finite spectral width via the
Fourier-transform relation between time and energy (set by the pulse width), and with a
finite angular spread, which is Fourier-related to the beam waist (set by the focusing
optics).
In contrast, the natural ALP decay rate,
\(\Gamma = (g/M)^2 m_a^{3} / (16\pi)\), that enters
the Breit–Wigner propagator is extremely small in the coupling range of interest. 
For example, for \(g/M = 10^{-10}\,\mathrm{GeV}^{-1}\) and \(m_a = 1\,\mathrm{eV}\),
one finds \(\Gamma \simeq 2 \times 10^{-40}\,\mathrm{eV}\).
Compared to any realistic experimental window \(\Delta E_{\mathrm{cms}}\) on \(E_{\mathrm{cms}}\),
the Breit-Wigner width is therefore almost delta-function-like.
However, the integrated effect of the s-channel pole, including the off-shell part of
the Breit-Wigner function within \(\Delta E_{\mathrm{cms}}\), can drastically increase
the sensitivity~\cite{JHEP}.
In our scheme, the observable lineshape is in practice governed by the experimental
distribution of \(E_{\mathrm{cms}}\), rather than by the tiny intrinsic ALP width.
Given an accurately controlled \(\Delta E_{\mathrm{cms}}\) in the experiment, if an ALP
resonant mass lies within that range, we can claim the existence of a resonance
through the averaged pole contribution.

Experimentally, the overlap factor $\mathcal{D}_{three}$ is expected to deteriorate due to variations in the incidence angle of each beam, $\Delta{\theta_{x,j}}, \Delta{\theta_{y,j}},(j=c1,c2,i)$, which in turn causes shifts in peak positions at the focal point, $\Delta{x}_j, \Delta{y}_j, c\Delta{t}_j ,(j=c1,c2,i)$.
Therefore, in the search, the peak positions of the focal spot profiles of the three laser beams were measured before and after each run r = 1,2,3 using IPM2. The focal position measured before each of the three runs was treated as the ideal overlap factor with zero misalignment, denoted as $\mathcal{D}_{0}$. The overlap factor after each run including the three beam offsets is expressed using the measured parameters as $\mathcal{D}_{r}(\Delta{x}_j,\Delta{y}_j,c\Delta{t}_j,\Delta{\theta_{x,j}},\Delta{\theta_{y,j}})$. Here, the reduction in the overlap factor from timing and angular mismatches $c\Delta{t}_j,\Delta{\theta_{x,j}},\Delta{\theta_{y,j}}$ is much smaller than that from the transverse offsets $\Delta{x}_j,\Delta{y}_j$. We therefore retain only the $\Delta x_j,\ \Delta y_j$ contributions. Finally, the merged and averaged overlap factor $\overline{\mathcal{D}}_{\mathrm{three}}$, is defined from the pre- and post-run values as

\begin{equation} \label{dfactor_compact}
\overline{\mathcal{D}}_{\mathrm{three}}
= \frac{1}{N_{\text{run}}}
\sum_{r=1}^{N_{\text{run}}}
\frac{\mathcal{D}_{0} + \mathcal{D}_{r}(\Delta{x}_j,\Delta{y}_j)}{2},
\end{equation}
with $N_{\text{run}} = 3$ in this experiment. Typical transverse shifts were $\Delta x_j,\Delta y_j \simeq \pm 3\,\mu\mathrm{m}$ for $j\in\{c1,c2,i\}$. For reference, the measured waist diameters at the CP were $2w_c \simeq 30\,\mu\mathrm{m}$ (creation beams) and $2w_i \simeq 52\,\mu\mathrm{m}$ (inducing beam). Further details of the three-beam overlap factor, $\overline{\mathcal{D}}_{\text{three}}$, are given in Appendix~\ref{App.C}.
Therefore, the experimental yield $\mathcal{Y}_{c+i}$ in Eq.~\eqref{yield0} is re-expressed as
        
        \begin{equation}
        \label{yield1}
        \mathcal{Y}_{c+i} \equiv N_{1} N_{2} N_{4} \overline{\mathcal{D}}_{\mathrm{three}} \left[s / L^3\right] \bar{\Sigma}_I \left[L^3 / s\right].
        \end{equation}

Based on the set of experimental parameters $P$ summarized in Table \ref{parameter},
the expected number of signal photons mediated by an ALP of mass $m_{a}$ and the coupling $g /M$ 
to two photons is expressed as
        \begin{equation}  \label{obtain_photon}
        n_{obs} = \mathcal{Y}_{c+i} \left(m_{a}, g / M; P\right) N_{shot} \epsilon,
        \end{equation}
where $N_{shot}$ is the number of shots in S-pattern and $\epsilon$ is 
the overall detection efficiency.
A coupling constant $g /{M}$ can be evaluated by numerically solving Eq.\eqref{obtain_photon} 
for an ALP mass $m_{a}$ and a given observed number of photons $n_{obs}$.

\begin{table*}[t]
  \caption{Experimental parameters used to obtain the upper limit.}
  \begin{center}
  \scalebox{1.00}[1.00] {
  \begin{tabular}{lr}  \\ \hline
  Parameter & Value \\ \hline
  Centeral wavelength of creation laser, $\lambda_{c}$   & 810.6 nm\\
  Relative linewidth of creation laser, $\delta\omega_{c}/<\omega_{c}>$ &  $1.5\times 10^{-2}$\\
  Duration time of creation laser (FWHM), $\sqrt{2\ln 2}~\tau_{c}$ & 34.2~fs \\
  Measured creation laser energy, $E_{c1}$ & 1.66 mJ \\
  Creation energy fraction within 3~$\sigma_{xy}$ focal spot, $\eta_{c1}$ & 0.89\\
  Effective creation energy within 3~$\sigma_{xy}$ focal spot & $E_{c1} \eta_{c1}$ = 1.48~mJ\\
  Effective number of creation photons, $N_{c1}$ & $ 6.0\times 10^{15} $ photons\\
  Beam diameter of creation laser beam, $d_{c1}$ & 5.0~mm\\
  Polarization, $\theta_{c1}$, $\epsilon_{c1}$ & $\theta_{c1}$ = 1.56~rad, $\epsilon_{c1}$ = 0.61~rad\\
  Measured creation laser energy, $E_{c2}$ & 1.26 mJ \\
  Creation energy fraction within 3~$\sigma_{xy}$ focal spot, $\eta_{c2}$ & 0.88\\
  Effective creation energy within 3~$\sigma_{xy}$ focal spot & $E_{c2} \eta_{c2}$ = 1.11~mJ\\
  Effective number of creation photons, $N_{c2}$ & $ 4.5\times 10^{15}$ photons\\
  Beam diameter of creation laser beam, $d_{c2}$ & 5.0~mm\\
  Creation focal length, $f_{c}$ & $f_{c} $= 101.6~mm \\
  Polarization, $\theta_{c2}$, $\epsilon_{c2}$ & $\theta_{c2}$= 0.04~rad, $\epsilon_{c2}$ = 0.50~rad\\ \hline
  Central wavelength of inducing laser, $\lambda_i$   & 1064~nm\\
  Relative linewidth of inducing laser, $\delta\omega_{i}/<\omega_{i}>$ &  $4.52\times 10^{-5}$\\
  Duration time of inducing laser beam (FWHM), $\sqrt{2\ln 2}~\tau_{ibeam}$ & 7.1~ns\\
  Measured inducing laser energy, $E_{i}$ &  60.55~mJ \\
  Linewidth-based duration time of inducing laser, $\tau_{i}/2$ & $\hbar/(2\delta\omega_{i})$ = 6.24~ps\\
  Inducing energy fraction within 3~$\sigma_{xy}$ focal spot, $\eta_i$ & 0.86\\
  Effective inducing energy per $\tau_i$ within 3~$\sigma_{xy}$ focal spot & $E_{i} (\tau_i/\tau_{ibeam}) \eta_i $= 107.8~$\mu$J\\
  Effective number of inducing photons, $N_i$ & $5.8\times 10^{14}$~photons\\
  Beam diameter of inducing laser beam, $d_{i}$ & 7.0~mm\\
  Inducing focal length, $f_{i}$      & $f_{i} = 203.2$~mm \\
  Polarization & circular (right-handed state from observer) \\ \hline
  PMT quantum efficiency, $\epsilon_{QE}$ & 8.61~\% \\
  Efficiency of optical path from IP to PMT, $\epsilon_{opt}$ & 23.37~\% \\
  Total detection efficiency, $\epsilon$  & 2.0~\% \\ \hline
  Total number of shots in trigger pattern S, $W_S$   &  14934~shots\\
  $\delta{n}_{obs}$ & 67\\
  \hline
  \end{tabular}
  } 
  \end{center}
  \label{parameter}
  \end{table*}

From Eq.~\eqref{signal_photon}, since no significant signal was observed in the present search, an exclusion region is established. As a natural null hypothesis, we consider Gaussian fluctuations around zero. The confidence level for this null hypothesis is defined as 
\begin{widetext}
\begin{equation}
\mathrm{C.L.}
= \frac{1}{\sqrt{2\pi}\,\sigma}\int_{-\infty}^{\mu+\delta}
\exp\!\left[-\frac{(x-\mu)^2}{2\sigma^2}\right]\,dx
= \frac{1}{2}
+ \frac{1}{\sqrt{2\pi}\,\sigma}\int_{0}^{\mu+\delta}
\exp\!\left[-\frac{(x-\mu)^2}{2\sigma^2}\right]\,dx ,
\label{confidencelevel}
\end{equation}
\end{widetext}
where $\mu$ is the expected value of $x$ under the hypothesis and $\sigma$ is the standard deviation. In the search, the expected value $x$ corresponds to the number of signal photons $n_{obs}$
and $\sigma$ is one-standard deviation $\delta n_{obs}$.
From Eq.~\eqref{signal_photon}, the null result implies $\mu=n_{obs}=0$, and the acceptance-uncorrected uncertainty $\delta n_{obs}$ around $n_{obs}=0$ is estimated as the root-mean-square (RMS) of all relevant error components,
        \begin{equation}
        \label{Delta}
         \delta n_{obs} = \sqrt{52.4^2 + 13.9^2 + 25.5^2 + 30.6^2} \simeq 67.
        \end{equation}
For a one-sided \(95\%\) C.L., Eq.~\eqref{confidencelevel} implies \(\delta/\sigma \simeq 1.64\); accordingly, we exclude \(x > \mu + \delta\)~\cite{SAPPHIRES02}.

The upper limit in the parameter plane $\left(m_{a}, g / M\right)$ was estimated by numerically solving 
Eq.\eqref{N_obs} with the uncertainty $\delta n_{obs}$ defined in \eqref{Delta}, using the set of experimental parameters $P$ summarized in Tab.\ref{parameter}
        \begin{equation}
        \label{N_obs}  
        1.64 \delta n_{obs} = \mathcal{Y}_{c+i}\left(m_{a}, g / M; P\right) N_{shot} \epsilon,
        \end{equation}
where $N_{shot}=14,934$ and
the overall efficiency $\epsilon \equiv \epsilon_{opt}\epsilon_{QE}$ with
the optical path acceptance from IP to PMT1, $\epsilon_{opt}$, and the photocathode quantum efficiency of PMT1, $\epsilon_{QE}$, were substituted.
$\epsilon_{opt}$ was estimated by aligning the c1 laser line with the signal line ($\theta_s \sim 34^\circ$) in the setup diagram in Fig.\ref{setup} and measuring the energy at both the collision point and PMT1 positions using a common CMOS camera detector with a continuous calibration laser (He:Ne, 633 nm). The efficiency was determined as the ratio of the measured energy at these two locations. The He:Ne beam diameter was chosen to reproduce the expected signal photon angular distribution $\Delta{\theta_s}$, which was inferred from the error propagation via energy-momentum conservation.
$\epsilon_{QE}$ was estimated in a PMT calibration using a pulsed laser (wavelength $660$~nm)
as the ratio of the number of photoelectrons recorded by PMT1 to the number of incident photons, with the average photons per pulse held constant.

\begin{figure*}[t]
        \centering
          \includegraphics[width=0.78\linewidth]{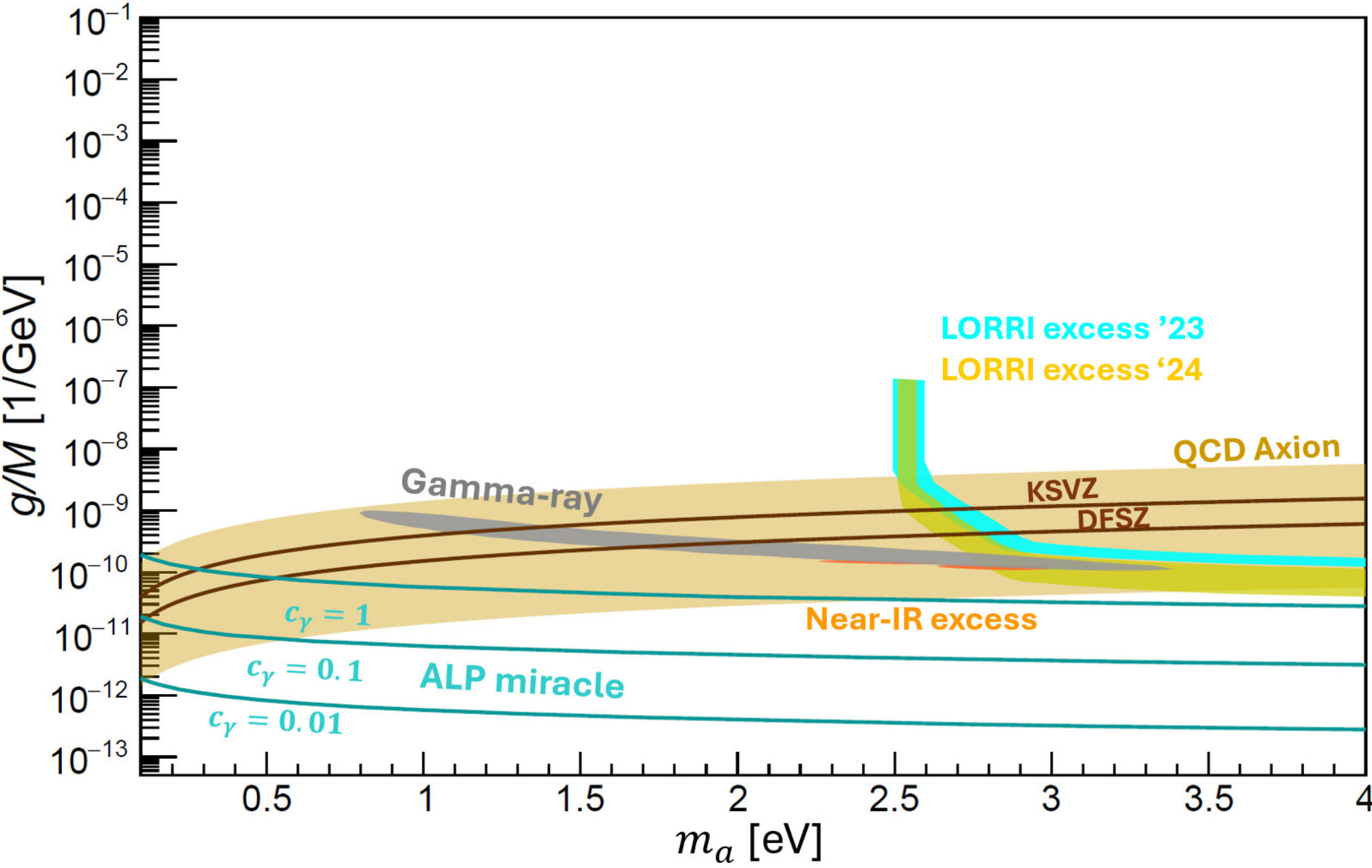}
  \medskip
  \includegraphics[width=0.78\linewidth]{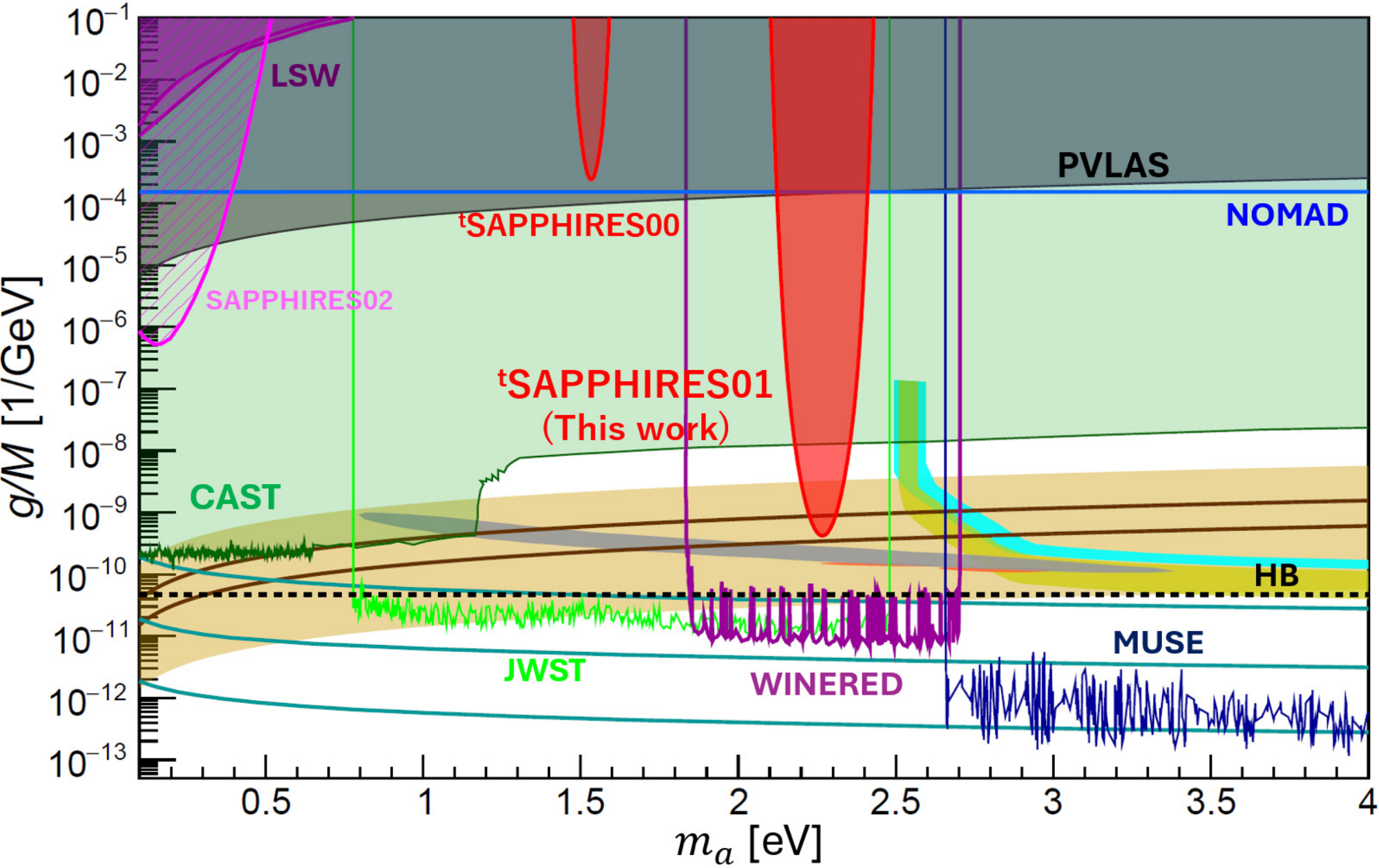}
\caption{
\textbf{Top:}
ALP-favored regions from COB/CIB and $\gamma$-ray attenuation are shown; the gray region shows blazar $\gamma$-ray attenuation (ALP-decay model) \cite{gamma}, the cyan and yellow bands show COB excess (New Horizons/LORRI '23,'24) \cite{LORRI2}, orange region shows near-IR anisotropies (HST+Spitzer; ALP+IHL fit) \cite{HST}.
The beige band and the upper brown solid line correspond to QCD-axion predictions in the KSVZ framework~\cite{KSVZ}, for
$0.07 < \left|E/N - 1.95\right| < 7$ and $E/N=0$, respectively.
The lower brown solid line shows the DFSZ prediction~\cite{DFSZ} with $E/N=8/3$.
The cyan lines indicate the ALP {\it miracle} model~\cite{MIRACLE2} 
for $c_{\gamma}=1, 0.1, 0.01$.
\textbf{Bottom:} The red solid contour encloses the 95\% confidence level upper-limit region in the coupling-mass plane for pseudoscalar exchange (labelled ${}^\mathrm{t}$SAPPHIRES01). The previous search, ${}^{\mathrm{t}}$SAPPHIRES00, presents the excluded region
based on the runs conducted in air at $\theta_c = 30^\circ$ by the red solid curve \cite{3beam01}.
The magenta region shows the SRPC exclusion in a quasi-parallel two-beam configuration (SAPPHIRES02)~\cite{SAPPHIRES02}
The purple shaded regions show the excluded parameter space from Light-Shining-through-a-Wall (LSW) searches by ALPS~\cite{ALPS} and OSQAR~\cite{OSQAR}.
The gray region indicates the limit from the vacuum magnetic birefringence (VMB) measurement by PVLAS~\cite{PVLAS}.
The light-cyan horizontal solid line denotes the NOMAD upper limit from the SPS neutrino beam search~\cite{NOMAD}.
The green region shows the exclusion from the CAST helioscope~\cite{CAST}.
The horizontal dotted line denotes the upper bound from the HB observation~\cite{HB2022,HB2014}.
The light-green regions indicate exclusions from JWST/NIRSpec adopting NFW profile of a Milky Way DM halo~\cite{JWST}.
The purple region indicates exclusion by WINERED spectroscopy adopting NFW profile of dSphs (Leo V, Tucana II); localized $>5\sigma$ excesses remain unexplained~\cite{WINERED}.
The blue region indicates exclusion by VLT/MUSE-Faint observations adopting NFW profile of a sample of five galaxies~\cite{MUSE}.
}
        \label{3beam01_all}
\end{figure*}

Figure~\ref{3beam01_all} discusses theoretically and phenomenologically favored domains and
existing upper limits in the ALP coupling-mass plane. For clarity, we first present the ALP-favored regions and theoretical benchmarks, and then overlay previously explored parameter spaces. 
 
Figure \ref{3beam01_all} (top) shows the ALP-favored regions and theoretical models: 
the gray region shows the parameter space favored by $\gamma$-ray attenuation of blazar spectra when the optical-depth excess is modeled as ALP two-photon decays \cite{gamma}.
The cyan and yellow regions show the parameter spaces favored by COB excess (New Horizons/LORRI ’23,’24) \cite{LORRI2}.
The orange regions show the parameter spaces favored by HST + Spitzer near-IR anisotropy data when fit with an ALP + intra-halo-light \cite{HST}. 
The beige shaded band corresponds to the benchmark QCD-axion model (KSVZ \cite{QCDband1,QCDband2}) with $0.07 < \left|E/N - 1.95\right| < 7$. The upper brown solid line shows the KSVZ model \cite{KSVZ} with $E/N=0$, while the lower brown solid curve indicates the DFSZ model \cite{DFSZ} with $E/N=8/3$.
The cyan lines illustrate the predictions of the ALP {\it miracle} scenario \cite{MIRACLE2} with the intrinsic parameters $c_{\gamma}=1, 0.1, 0.01$.

Figure \ref{3beam01_all} (bottom) shows
the upper limit in the coupling-mass plane from this search, ${}^\mathrm{t}$SAPPHIRES01, 
enclosed by the red solid curve. The limit was set at a 95\% confidence level for pseudoscalar-type ALP exchanges.
The peak sensitivity was at $m_a=2.27~\mathrm{eV}$ with $g/M \sim 4.2\times10^{-10}\,\mathrm{GeV}^{-1}$, using a central wavelength $\lambda_c=811~\mathrm{nm}$ and an incidence angle $\theta_c=47.9^\circ$.
This represents the first model-independent exclusion/sensitivity reaching KSVZ line ($E/N=0$) \cite{KSVZ}. 
The mass broadening at a relaxed coupling of $g/M \sim 4.2\times10^{-9}\,\mathrm{GeV}^{-1}$ was approximately $m_a=2.27 \pm 0.07~\mathrm{eV}$. The sensitivity spread arises from fluctuations in the center-of-mass system energy, driven by the finite momentum spread and bandwidth of the focused, short-pulse lasers \cite{3beam00}.
The previous search ${}^{\mathrm{t}}$SAPPHIRES00 presents the excluded region based the runs conducted in air at an incidence angle $\theta_c=30^\circ$ (corresponding to $m_a=1.54,\mathrm{eV}$) \cite{3beam01}.
We note that the label ${}^{\mathrm{t}}$SRPC00 used in Ref.\cite{3beam01} has been relabeled as ${}^{\mathrm{t}}$SAPPHIRES00 following reanalysis, to ensure consistency with ${}^{\mathrm{t}}$SAPPHIRES01
in the derivation of experimental parameters and the beam overlap factor.

The magenta shading shows the parameter space excluded by SRPC searches performed in a quasi-parallel collision setup (SAPPHIRES02)~\cite{SAPPHIRES02}.
The purple shaded regions correspond to constraints from Light-Shining-through-a-Wall (LSW) experiments, including results from ALPS~\cite{ALPS} and OSQAR~\cite{OSQAR}.
The gray region indicates the exclusion from the vacuum magnetic birefringence (VMB) measurement (PVLAS~\cite{PVLAS}).
The light-cyan horizontal solid line shows the upper bound obtained by NOMAD using the SPS neutrino beam~\cite{NOMAD}.
The green shaded region denotes the CAST helioscope exclusion~\cite{CAST}, obtained from searches for solar axions/ALPs via axion-photon conversion.
The horizontal dotted line is disfavoured by globular-cluster constraints based on the impact of ALP emission on stellar evolution \cite{HB2022,HB2014}.
The light-green regions indicate exclusions from JWST/NIRSpec-IFU blank-sky spectra~\cite{JWST}. These limits overlap our search region but depend on background modeling (smooth continuum model) and astrophysical assumptions about Milky Way DM halo (NFW), hence complementary.
The purple regions indicate exclusions from high-dispersion WINERED spectroscopy of dSphs (Leo V, Tucana II adopting NFW profile)~\cite{WINERED}. However, the authors also noted localized $>5\sigma$ excesses in that window, and its origin remains uncertain.
The blue regions indicate exclusions from the VLT/MUSE-Faint observations of a dSph (Leo T adopting NFW profile)~\cite{MUSE}.

\section{Conclusion}
We have presented the results of a single-point ALP search using a variable-angle three-beam stimulated resonant photon collider (\({}^\mathrm{t}\)SRPC). 
The energies of the creation lasers $c_1$ and $c_2$ (\(1.4~\mathrm{mJ}/34~\mathrm{fs}\), Ti:Sapphire) 
and the inducing laser i (\(60.6~\mathrm{mJ}/7.1~\mathrm{ns}\), Nd:YAG) were each increased 
by more than three orders of magnitude compared to the previous pilot search \cite{3beam01}. 
The experiment was performed in vacuum to suppress background photons. 
No statistically significant signal was observed. 
We therefore set exclusion limits, reaching a minimum coupling of 
\(g/M = 4.2 \times 10^{-10}~\mathrm{GeV}^{-1}\) at \(m_{a} = 2.27~\mathrm{eV}\), 
based on the formulation dedicated to \({}^\mathrm{t}\)SRPC~\cite{3beam00} and the experimentally
determined overlap factor presented in this work. 
This is the first model-independent upper limit that reaches the KSVZ benchmark in the eV range.
This single-point search demonstrates the feasibility of a smooth mass scan
by varying the collision angle in near future searches. 

\section*{Acknowledgments}
K. Homma acknowledges 
the support of the Collaborative Research Program of the Institute for Chemical Research at Kyoto University (Grant Nos. 2024-95 and 2025-100), the JSPS Core-to-Core Program (grant number: JPJSCCA20230003), and Grants-in-Aid for Scientific Research (Nos. and 24KK0068 and 25H00645) from the Ministry of Education, Culture, Sports, Science and Technology (MEXT) of Japan.
T. Hasada acknowledges support from the JST SPRING, Grant No. JPMJSP2132, and a Grant-in-Aid for JSPS fellows No. 25KJ1860 from the Ministry of Education, Culture, Sports, Science and Technology (MEXT) of Japan.
The $T^{6}$ system was financially supported by
the MEXT Quantum Leap Flagship Program (JPMXS0118070187) and
the program for advanced research equipment platforms (JPMXS0450300521).

\appendix
\setcounter{section}{0}   

\section{PMT calibration} \label{App.A}

In this measurement, we quantitatively evaluate single-photon signals with a photomultiplier tube (PMT). The photocathode quantum efficiency $\epsilon_{QE}$ is wavelength dependent, and the electron multiplication in the dynode chain is well described by a gamma distribution. We thus calibrated the one-photon-equivalent charge near the signal wavelength, using a pulsed white-light source (SuperK Compact; pulse width $t<1~\mathrm{ns}$) and a band-pass filter ($660\pm5~\mathrm{nm}$).

The baseline (zero-photon) integrated charge follows a Gaussian distribution $G_0$, while the charge distribution for $n\ge 1$ detected photon after the PMT follows a gamma distribution $g_n$. Denoting by $q$ the baseline-subtracted integrated charge, the probability densities are
\begin{align}
G_0\!\left(q\,\big|\,m_0,s_0\right)
&=\frac{1}{\sqrt{2\pi}\,s_0}\exp\!\left[-\frac{(q-m_0)^2}{2s_0^2}\right],\\[4pt]
g_n\!\left(q\,\big|\,n\kappa,\theta\right)
&=\frac{1}{\Gamma(n\kappa)\,\theta^{\,n\kappa}}\,
(-q)^{\,n\kappa-1}\exp\!\left(\frac{q}{\theta}\right)\qquad(q\le 0),
\label{A.1}
\end{align}
where, $m_0, s_0$ are the mean and standard deviation of the baseline, respectively; $\theta$ is the scale parameter and $\kappa$ is the shape parameter. Mean $m_1$ and standard deviation $s_1$
of the 1-photon charge distribution follow,
\begin{equation}
\kappa=\left(\frac{m_1}{s_1}\right)^{\!2}, \qquad
\theta=-\frac{s_1^{2}}{m_1}.
\label{A.2}
\end{equation}

On the other hand, the number of photons detected by PMT follows a Poisson distribution with mean $\lambda$. Therefore, the measured charge distribution can be modeled as a Poisson mixture of baseline (zero-photon) and $n$-photon response components:
\begin{equation}
F(q;\lambda)
= A\Bigg[
e^{-\lambda}G_0\!\left(q\,\big|\,m_0,s_0\right)
+\sum_{n=1}^{\infty}\frac{e^{-\lambda}\lambda^{n}}{n!}\;
g_n\!\left(q\,\big|\,n\kappa,\theta\right)
\Bigg].
\end{equation}
Here $A$ is an overall normalization (amplitude), $G_0$ and $g_n$ are unit-normalized probability density functions (PDFs) for the baseline and the $n$-photon response, respectively.

\begin{figure*}[t]
        \centering
        \includegraphics[width=1.00\textwidth]{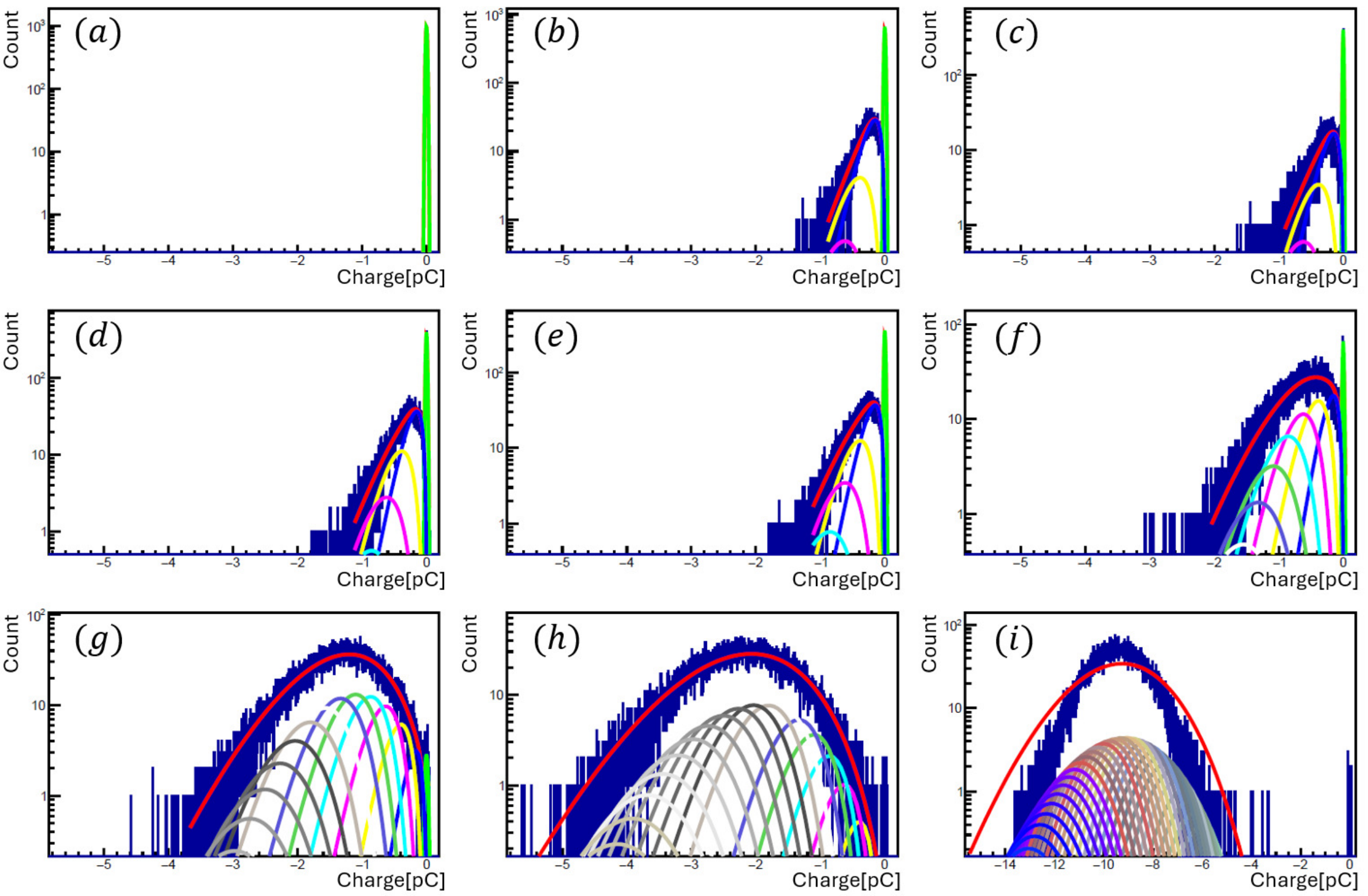}
        \caption{
        Several-photon charge distributions fitted with a Poisson-weighted mixture (red curve): a Gaussian pedestal for $0$-photon and Polya/Gamma components for $n\ge 1$ photons. Colored curves show the individual Poisson components $e^{-\lambda}\lambda^n/n!$ multiplied by the corresponding $n$-PE response $(a-h)$. With multiple photons, a nonlinear PMT response is visible in the lower-right panel $(i)$.
}
\label{PMT}
\end{figure*}

Figure~\ref{PMT} shows charge distributions (vertical axis: number of shots per charge bin) as the mean photon number $\lambda$ is stepped from the zero-photon level upward $(a-i)$. At high photon occupancy, the PMT exhibits clear nonlinearity, as seen in the lower-right panel $(i)$. At each photon-number level, we acquired $30{,}000$ shots. The charge distribution are produced by the threshold-based peak-finding algorithm described in Sec. \ref{Sec4}. For shots that do not cross the threshold $L_{5\sigma}$, we compute the integrated charge over a fixed 5~ns “typical-waveform” window for 0-photon distribution.

The fitting proceeds as follows:
\begin{enumerate}
\item[(i)] Fit the zero-photon distribution (panel~(a)) with a Gaussian to extract $(m_0,s_0)$.
\item[(ii)] Fit the low-occupancy dataset (panel~(b)) with $(m_0,s_0)$ held fixed to determine the one-photon parameters $(m_1,s_1)$ of the gamma response.
\item[(iii)] For the higher-occupancy datasets (panels~(c)-(h)), fit with $(m_0,s_0,m_1,s_1)$ fixed and vary $\lambda$ and normalization $A$ only.
\end{enumerate}

From this procedure we obtain a linear relation between the photon expectation value $\lambda$ and the mean (baseline-subtracted) charge,
%
\begin{equation}
Q(\lambda)=Q_{1PE}\,\lambda + b. 
\end{equation}

\begin{figure}[!htbp]
        \centering 
        \includegraphics[width=0.48\textwidth]{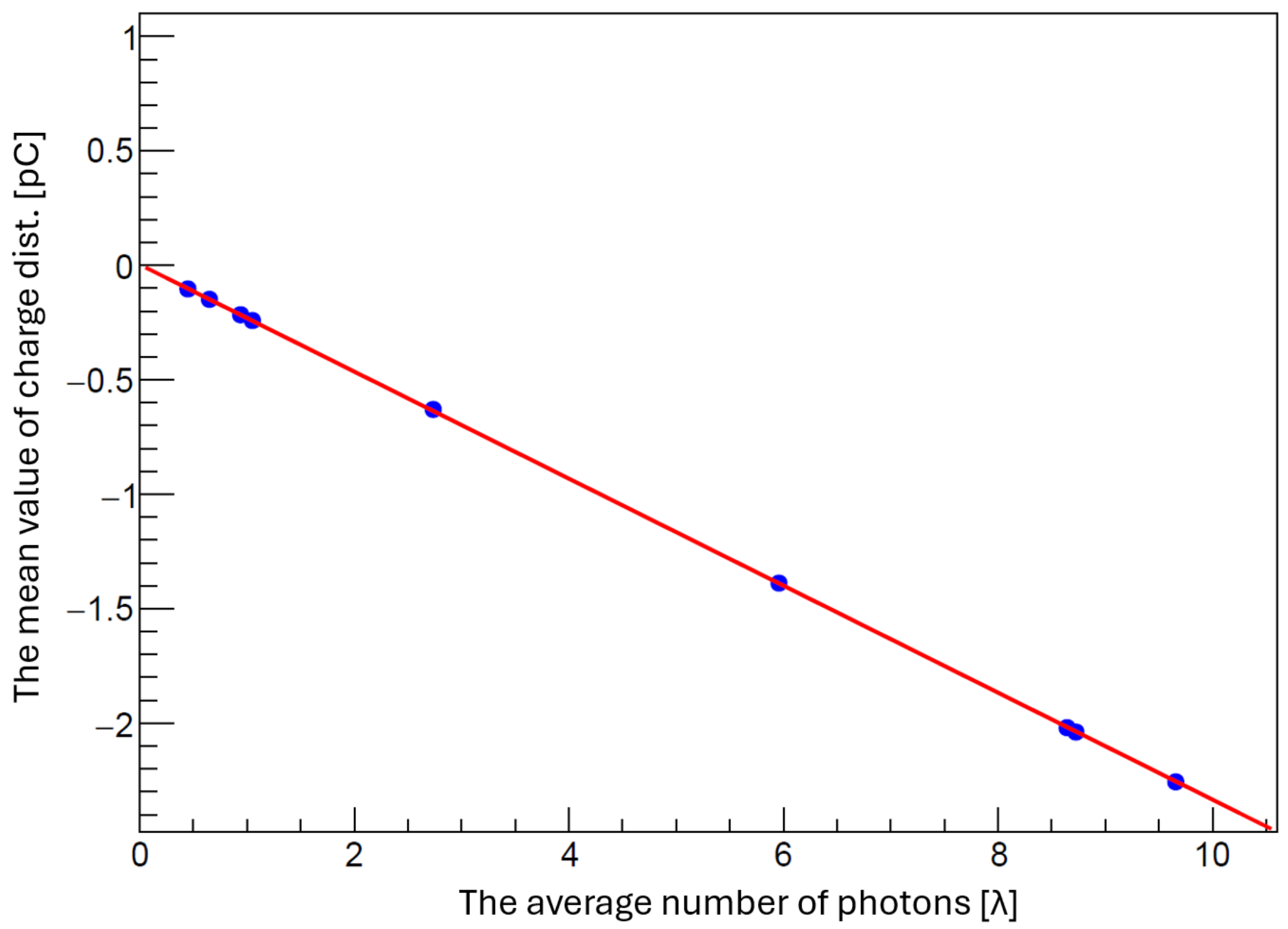}
        \caption{ 
        Linear relation between the fit-averaged number of photoelectrons and the mean charges. The blue dots show the measurements in Fig. \ref{PMT} (b)-(h) plus additional datasets not shown in Fig. \ref{PMT}. The linear fit is performed for data with $\lambda<10$, and the slope corresponds to the one-photon equivalent charge.
}
\label{1PEQ}
\end{figure}

Figure \ref{1PEQ} shows the mean value of charge distributions $Q$ versus photon expectation value $\lambda$ ($\lambda<10$) together with the best-fit line. The slope parameter $Q_{1PE}$ equals the one-photon-equivalent charge.

\section{Statistical uncertainty from Gamma-distributed single-photon charge fluctuations} \label{App.B}
In this appendix we derive the statistical uncertainty of the photon-number estimator used in the analysis, starting from the fact that the photomultiplier tube (PMT) records analog integrated charges rather than integer photon counts. The estimator is constructed from the measured charge using the calibrated one-photon-equivalent charge, and its variance receives contributions both from Poisson photon statistics and from the finite width of the single-photon charge response.

For clarity, we first consider the idealized case in which the photon number is fixed to a particular integer value $n$ and does not fluctuate from trigger to trigger. In that case the total charge can be written as
\begin{equation}
  Q_{n} = \sum_{j=1}^{n} q_{j},
\end{equation}
where $q_{j}$ denotes the charge contributed by the $j$-th detected photon ($j = 1,\dots,n$). Each $q_{j}$ is an independent and identically distributed random variable corresponding to the single-photon Gamma response (the $n=1$ component of Eq.~\eqref{A.1}), with mean $m_{1}$ and variance $s_{1}^{2}$. We use $m_{1}$ as the one-photon-equivalent charge. In this fixed-$n$ ensemble one has
\begin{equation}
  \langle Q_{n} \rangle = n\,m_{1}, \qquad \mathrm{Var}(Q_{n}) = n\,s_{1}^{2}.
\end{equation}

In the actual experiment, however, the photon number $n$ fluctuates from trigger to trigger according to the Poisson law with mean $\lambda$, so that
\begin{equation}
  \langle n \rangle = \lambda, \qquad \mathrm{Var}(n) = \lambda.
\end{equation}

The total variance of the measured charge $Q$ can then be decomposed, according to the law of total variance, as
\begin{equation}
\begin{aligned}
  \mathrm{Var}(Q)
  &= \big\langle \mathrm{Var}(Q_{n}) \big\rangle
  + \mathrm{Var}\big( \langle Q_{n} \rangle \big) \\
  &= \big\langle n s_{1}^{2} \big\rangle + \mathrm{Var}(n m_{1}) \\
  &= s_{1}^{2} \langle n \rangle + m_{1}^{2} \mathrm{Var}(n) \\
  &= \lambda s_{1}^{2} + \lambda m_{1}^{2} \\
  &= \lambda m_{1}^{2}\left( 1 + \frac{s_{1}^{2}}{m_{1}^{2}} \right).
\end{aligned}
\end{equation}

It is convenient to define an effective Fano factor directly for the charge variable $Q$ as
\begin{equation}
  F \equiv \frac{\mathrm{Var}(Q)}{m_{1}\,\langle Q\rangle}
  = 1 + \frac{s_{1}^{2}}{m_{1}^{2}}
  = 1 + \frac{1}{\kappa},
\end{equation}
where we used $\kappa = (m_{1}/s_{1})^{2}$ from Eq.~\eqref{A.2}. For the measured single-photon response we obtain $\kappa \simeq 2.81$, corresponding to
\begin{equation}
  F \simeq 1.36.
\end{equation}
Thus the charge variance scales as
\begin{equation}
  \mathrm{Var}(Q) = F\,m_{1}^{2}\,\lambda.
\end{equation}

In practice, the Poisson mean $\lambda$ is not known a priori and is estimated from the data. Defining the photon-number estimator by
\begin{equation}
  N = \frac{Q}{m_{1}},
\end{equation}
we have $\langle N \rangle = \lambda$, so that $N$ is an unbiased estimator of the mean photon number. In the following we therefore replace $\lambda$ by $N$ and use
\begin{equation}
  \mathrm{Var}(Q) \simeq F\,m_{1}^{2}\,N
\end{equation}
in the error propagation.

The one-photon-equivalent charge $Q_{1PE} \equiv m_{1}$ is known with a finite calibration uncertainty $\delta Q_{1PE}$ from Eq.~\eqref{Q1PE}. Treating $Q$ and $Q_{1PE}$ as independent quantities, the standard error propagation for
\begin{equation}
  N = \frac{Q}{Q_{1PE}}
\end{equation}
gives
\begin{eqnarray}
(\delta N)^{2}
&=& \left(\frac{\partial N}{\partial Q}\right)^{2} (\delta Q)^{2}
+ \left(\frac{\partial N}{\partial Q_{1PE}}\right)^{2} (\delta Q_{1PE})^{2} \nonumber \\
&=& \frac{\mathrm{Var}(Q)}{Q_{1PE}^{2}}
+ N^{2} \left(\frac{\delta Q_{1PE}}{Q_{1PE}}\right)^{2}.
\end{eqnarray}
Using $\mathrm{Var}(Q) = F Q_{1PE}^{2} N$, we obtain
\begin{equation}
(\delta N)^{2}
= F N + N^{2} \left(\frac{\delta Q_{1PE}}{Q_{1PE}}\right)^{2}.
\end{equation}

The first term represents the genuine statistical uncertainty of the photon-number estimator, while the second term, induced by the finite precision of the calibration of $Q_{1PE}$, is treated as a global calibration systematic. For the present data set, this second term evaluates to $\sim 3 \times 10^{-3}$, which is negligible compared to the leading statistical term and is therefore ignored in the following.

For each trigger pattern $X \in \{S, C, I, P\}$, the reconstructed photon number $N_{X}$ follows
\begin{equation}
  \mathrm{Var}(N_{X}) \simeq F\,N_{X}.
\end{equation}
The signal photon yield per shot is obtained from the linear combination
\begin{equation}
  n_{\mathrm{sig}} = N_{S} - N_{C} - N_{I} + N_{P},
\end{equation}
so that, neglecting correlations between different patterns, the corresponding statistical variance is
\begin{equation}
  \delta n_{\mathrm{sig}}^{2}
  = F \left( N_{S} + N_{C} + N_{I} + N_{P} \right).
  \label{stat}
\end{equation}
This expression is used to evaluate the statistical error quoted in the main text.

\section{3-beam overlap factor $\overline{\mathcal{D}}_{\mathrm{three}}$ including offset} \label{App.C}
During data taking, we need to enforce spatial overlap at the collision point (CP) and temporal overlap between the two femtosecond creation pulses. The spatial condition is ensured by aligning the peak of each laser profile to the single camera pixel calibrated with the center of a crossed-wire target, while the temporal condition is set by maximizing the second-harmonic (SHG) signal from a thin nonlinear crystal (NLC) during a delay-stage scan. We define this pre-run condition as the ideal overlap factor.
However, focal drifts and time shifts inevitably accumulate over a $\sim 30\,\mathrm{min}$ run, causing the actual overlap to deviate from the ideal. These misalignments can be attributed to beam-pointing (angular) drifts. In practice, refractive-index fluctuations in upstream crystals driven by temperature modulation can seed those deviations. The transverse focal-spot shifts of each beam are measured with the collision-point monitor (CPM2), and incidence-angle deviations and timing offsets at the CP are inferred by back-propagating these shifts using the paraxial ray-transfer (ABCD) matrix formalism \cite{Yariv}. In our previous study~\cite{GHzmixing}, we introduced an overlap factor that included only focal offsets. In this work, we generalized that factor to incorporate ABCD-inferred incidence-angle deviations and timing offsets, and then estimated their contributions. In our setup, however, these effects proved numerically subdominant to transverse position offsets at the CP.
Accordingly, this appendix introduces a simplified three-beam overlap factor that retains only the focal-drift terms of Ref.~\cite{GHzmixing}. We quantify the impact of this reduction below and define the effective overlap factor $\overline{\mathcal{D}}_{\mathrm{three}}$ used in the present analysis.

\begin{figure*}[t]
        \centering
        \includegraphics[width=1.00\textwidth]{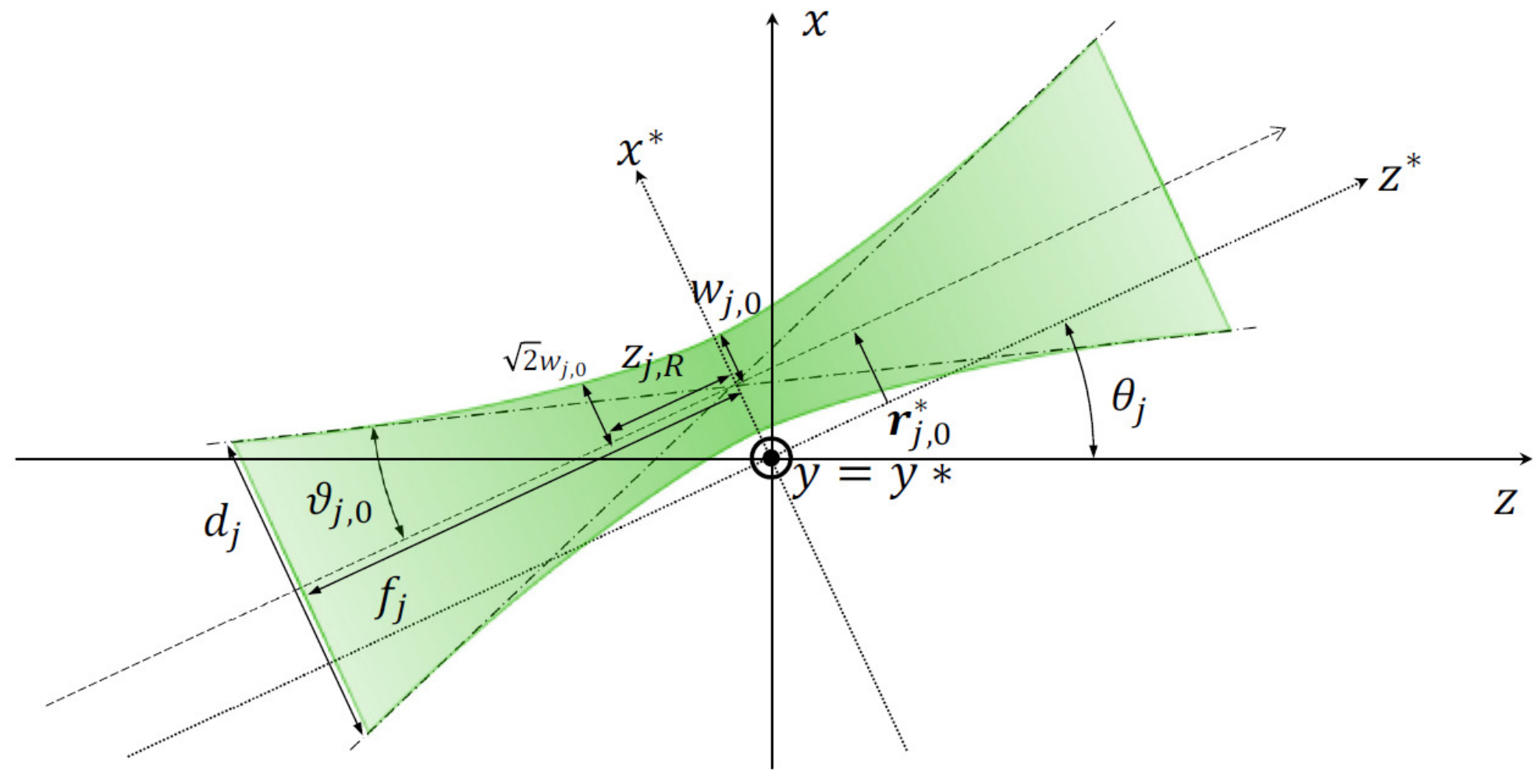}
        \caption{
Geometry and parameters for an each focused beam at incidence angle $\theta_j$ ($j=c_{1},c_{2},i$). Beam-attached coordinates
(${}^{*}$) are defined by rotating the lab frame such that the beam propagates along $z^{*}$. 
The transverse offset at focus is $\mathbf{r}^{*}_{j,0}$ in the $x^{*}-y^{*}$ plane.
The focusing half-angle is $\vartheta_{j,0}=\arctan\!\bigl(d_j/(2f_j)\bigr)$, where $d_j$ and $f_j$ are the beam diameter and focal length.
The waist at focus is $w_{j,0}$. Along $z^{*}$, the radius increases with distance and reaches $w_j=\sqrt{2}\,w_{j,0}$ at the Rayleigh length $z_{j,R}$. Adapted from Ref.~\cite{GHzmixing}.
}
\label{BeamGeo}
\end{figure*}

As illustrated in Figure \ref{BeamGeo}, we introduce a set of beam-geometry parameters that allow for non-zero transverse drifts \cite{GHzmixing}. The laser-beam coordinate system rotated by $\theta_j$ ($j=c_{1},c_{2},i$) is denoted with the asterisk symbol.
Propagation is along the dashed arrow (parallel to $z^{*}$), and its displacement $\mathbf{r}^{*}_{j,0}$ is defined in the $x^{*}\!-\!y^{*}$ plane. The experimentally extended overlap factor is parameterized as

\begin{widetext}
\begin{align}
\mathcal{D}_{\exp}
&= \left(\frac{2}{\pi}\right)^{3/2} \, w_{i,0}^{2}\, c_{0}\tau_{i}
\int_{-z_{i,R}/c_{0}}^{+z_{i,R}/c_{0}}\! dt\;
\frac{\displaystyle \prod_{j}\!\bigl(w_{j}^{-2}(c_{0}\tau_{j})^{-1}\bigr)}
{\displaystyle \sqrt{\sum_{j} w_{j}^{-2}}\;
\Bigl(\sum_{j}\delta_{j}\Bigr)\!
\Bigl(\sum_{j}(\mu_{j}^{2}+\nu_{j}^{2})\Bigr)}
\notag\\[2pt]
&\quad\times
\exp\!\biggl[
-2\biggl\{
\sum_{j}\!\bigl(\xi_{j}^{2}+\eta_{j}^{2}+\zeta_{j}^{2}\bigr)
-
\frac{\Bigl(\sum_{j}(\mu_{j}\xi_{j}+\nu_{j}\zeta_{j})\Bigr)^{2}}
{\sum_{j}(\mu_{j}^{2}+\nu_{j}^{2})}
\biggr\}
\biggr],
\end{align}
with the following parameter definitions

\[
\begin{aligned}
\alpha_j &\equiv \frac{\cos\theta_j}{w_j^{2}},\qquad
\beta_j  \equiv \frac{\sin\theta_j}{c_0^{2}\,\tau_j^{2}},\qquad
\delta_j \equiv \alpha_j \cos\theta_j + \beta_j \sin\theta_j,\\[4pt]
\mu_j &\equiv
\frac{1}{\sqrt{\sum_{m}\delta_m}}\;
\sum_{k,l}\varepsilon_{jkl}\,
\sqrt{\frac{\delta_l}{\delta_k}}\,
\bigl(\alpha_k \sin\theta_k - \beta_k \cos\theta_k\bigr),\\[4pt]
\nu_j &\equiv \frac{1}{w_j\,c_0\,\tau_j}\,\sqrt{\delta_j},\\[4pt]
\xi_j &\equiv
\frac{1}{\sqrt{\sum_{m}\delta_m}}\;
\sum_{k,l}\varepsilon_{jkl}\,
\sqrt{\frac{\delta_l}{\delta_k}}\,
\bigl(\alpha_k x^{*}_{k,0} + \beta_k c_0 t_k\bigr),\\[4pt]
\eta_j &\equiv
\frac{1}{\sqrt{\sum_{m} w_m^{-2}}}\;
\sum_{k,l}\varepsilon_{jkl}\,w_k^{-1}w_l^{-1}\,y^{*}_{k,0},\\[4pt]
\zeta_j &\equiv
\frac{1}{w_j\,c_0\,\tau_j}\,\sqrt{\delta_j}\,
\bigl(x^{*}_{j,0}\sin\theta_j - c_0 t_j \cos\theta_j\bigr)\,.
\end{aligned}
\]
\end{widetext}

The quantities \(\xi_j\), \(\eta_j\), and \(\zeta_j\) arise from drifts in the propagation directions of the individual pulses within the \(x^{*}\!-\!y^{*}\) plane, parameterized by \(x^{*}_{j,0},\,y^{*}_{j,0} \). In particular, \(\xi_j\) and \(\zeta_j\) depend explicitly on \(x^{*}_{j,0}\) and \(c_{0}t_j\) because rotations of the pulse trajectories in the \(z\!-\!x\) plane couple the \(x\) and \(z\) components. 
In Ref.~\cite{GHzmixing}, the time-integration window was set from one Rayleigh length upstream during the focusing of the inducing beam to the CP, \(t\in[-z_{i,R}/c_{0},\,0]\), as a conservative choice. When beam misalignments (transverse position) are included, some genuine instances of temporal overlap may fall outside this one-sided window and thus be missed. To estimate the reduction of the overlap factor due to misalignments relative to the ideal case in a symmetric manner, we adopt the symmetric interval \(t\in[-z_{i,R}/c_{0},\, +z_{i,R}/c_{0}]\) in this work.


Experimentally, we evaluate focal drifts by comparing the pre- and post-run peak positions of the focal profiles for each beam measured with CPM2. Because the CPM2 axes coincide with the beam-attached axes $(x^{*},y^{*})$, we henceforth denote $(x^{*}_{j,0},y^{*}_{j,0})\equiv(\Delta x_{j},\Delta y_{j})$ on the camera plane and drop the asterisks. 
We estimate the overlap for each run \(r\) by averaging the pre-run ideal value \(\mathcal{D}_0\) and the post-run value $\mathcal{D}_r(\Delta x_{j},\Delta y_{j})$:
\begin{equation}
\overline{\mathcal{D}}^{(r)} \;=\; \tfrac{1}{2}\,\bigl(\mathcal{D}_0 + \mathcal{D}_r\bigr).
\end{equation}
Table~\ref{drift} summarizes the run-by-run transverse offsets for each laser and the corresponding ideal/post-run overlap values.
Finally, the overall overlap $\overline{\mathcal{D}}_{\mathrm{three}}$used in the analysis is the run-averaged value
\begin{equation}
\overline{\mathcal{D}}_{\mathrm{three}}
\;=\;
\frac{1}{N_{\mathrm{run}}}\sum_{r=1}^{N_{\mathrm{run}}}\overline{\mathcal{D}}^{(r)}.
\end{equation}
For the runs listed in Table~\ref{drift}, this procedure yields \(\overline{\mathcal{D}}_{\mathrm{three}} \sim 5.09\).

\begin{table}[!htbp] 
  \centering
  \caption{Run-by-run focal-plane offsets [$\mu$m], and the corresponding ideal/post-run overlap values [$s / L^3$]}    \label{drift}
  \begin{tabular}{lrrrrrrrr}
    \hline
    & $\Delta x_{c1}$ & $\Delta y_{c1}$ & $\Delta x_{c2}$ & $\Delta y_{c2}$ & $\Delta x_{i}$ & $\Delta y_{i}$ & $\mathcal{D}_0$ & $\mathcal{D}_r$ \\
    \hline
    Run01 & -2.4 & -4.9 & -3.7 & -5.7 &  3.1 & 4.0 & 5.90 & 2.93\\
    Run02 & -4.5 & -0.7 &  1.8 & -0.4 &  0.0 & 0.6 & 5.90 & 5.32\\
    Run03 & -3.4 & -3.9 &  2.1 &  1.1 &  2.9 & 0.5 & 5.90 & 4.58\\
    \hline
  \end{tabular}
\end{table}

\end{document}